%% file: paper.tex
\documentclass[journal]{vgtc}                

\ifpdf
  \pdfoutput=1\relax                   
  \usepackage{graphicx}                
  \DeclareGraphicsExtensions{.pdf,.png,.jpg,.jpeg} 
\else
  \usepackage{graphicx}                
  \DeclareGraphicsExtensions{.eps}     
\fi%

\graphicspath{{figures/}{pictures/}{images/}{./}} 

\usepackage{microtype}                 
\PassOptionsToPackage{warn}{textcomp}  
\usepackage{textcomp}                  
\usepackage{mathptmx}                  
\usepackage{times}                     
\usepackage{cite}                      
\usepackage{tabu}                      
\usepackage{booktabs}                  
\usepackage{enumitem}
\usepackage{float}
\usepackage{amsmath}
\usepackage{xcolor}
\usepackage{setspace}

\definecolor{tblue}{HTML}{4472C4}
\definecolor{tgreen}{HTML}{70AD47}
\definecolor{torange}{HTML}{ED7D31}

\definecolor{falseadvertising}{HTML}{fb7161}
\definecolor{protectedproducts}{HTML}{418abd}
\definecolor{inappropriatebusiness}{HTML}{fc9c30}
\definecolor{sensitivecontent}{HTML}{50bba9}
\newcommand{\ttfa}[1]{\textbf{\textcolor{falseadvertising}{#1}}}
\newcommand{\ttpp}[1]{\textbf{\textcolor{protectedproducts}{#1}}}
\newcommand{\ttib}[1]{\textbf{\textcolor{inappropriatebusiness}{#1}}}
\newcommand{\ttsc}[1]{\textbf{\textcolor{sensitivecontent}{#1}}}

\newcommand{\trevise}[1]{\textcolor{black}{#1}}

\onlineid{1218}

\vgtccategory{Research}
\vgtcpapertype{application/design study}

\title{VideoModerator: A Risk-aware Framework for Multimodal Video Moderation in E-Commerce}

\author{Tan Tang, Yanhong Wu, Lingyun Yu, Yuhong Li and Yingcai Wu}
\authorfooter{
\item
 Tan Tang, Yanhong Wu and Yingcai Wu are with State Key Lab of CAD\&CG, Zhejiang University and Zhejiang Lab. E-mail: tangtan, yanhongwu, ycwu@zju.edu.cn. Yingcai Wu is the corresponding author.
\item
 Lingyun Yu is with Department of Computing, Xi’an Jiaotong-Liverpool University. E-mail: Lingyun.Yu@xjtlu.edu.cn.
\item
 Yuhong Li is with Alibaba Group. E-mail: daniel.lyh@alibaba-inc.com.
}

\shortauthortitle{Tang \MakeLowercase{\textit{et al.}}: VideoModerator}

\abstract{
Video moderation, which refers to remove deviant or explicit content from e-commerce livestreams, has become prevalent owing to social and engaging features.
However, this task is tedious and time consuming due to the difficulties associated with watching and reviewing multimodal video content, including video frames and audio clips. 
To ensure effective video moderation, we propose VideoModerator, a risk-aware framework that seamlessly integrates human knowledge with machine insights.
This framework incorporates a set of advanced machine learning models to extract the risk-aware features from multimodal video content and discover potentially deviant videos. 
Moreover, this framework introduces an interactive visualization interface with three views, namely, a video view, a frame view, and an audio view.
In the video view, we adopt a segmented timeline and highlight high-risk periods that may contain deviant information.
In the frame view, we present a novel visual summarization method that combines risk-aware features and video context to enable quick video navigation. 
In the audio view, we employ a storyline-based design to provide a multi-faceted overview which can be used to explore audio content.
Furthermore, we report the usage of VideoModerator through a case scenario and conduct experiments and a controlled user study to validate its effectiveness.
} 


\CCScatlist{ 
 \CCScat{K.6.1}{Management of Computing and Information Systems}%
{Project and People Management}{Life Cycle};
 \CCScat{K.7.m}{The Computing Profession}{Miscellaneous}{Ethics}
}

\vgtcinsertpkg

\begin{document}
\begin{spacing}{0.94}

\firstsection{Introduction}

\maketitle

\input{sec1_intro.tex}
\input{sec2_related.tex}
\input{sec3_framework.tex}
\input{sec4_visualdesign.tex}
\input{sec5_cases.tex}
\input{sec6_userstudy.tex}
\input{sec7_conclusion.tex}

\acknowledgments{
The work was supported by NSFC (62072400) and Zhejiang Provincial Natural Science Foundation (LR18F020001).
This work was also supported by Alibaba-Zhejiang University Joint Institute of Frontier Technologies and the Collaborative Innovation Center of Artificial Intelligence by MOE and Zhejiang Provincial Government (ZJU).
}

\newpage

\bibliographystyle{abbrv-doi}

\bibliography{reference}
\end{spacing}
\end{document}

%% file: sec1_intro.tex
E-commerce livestreaming, in which a vendor is a livestreamer who advertises and promotes his/her products on e-commercial platforms~\cite{Cai2019a,Sun2019a}, has taken off.
Such online commercial activities generate a large amount of user-generated videos that may include misinformation or offensive content, which can exert a negative or damaging impact on e-commerce platforms~\cite{Roberts2016a}.
To protect platforms' business and attract consumers by using innovative content, video moderation, which involves reviewing videos to remove unnecessary or explicit content~\cite{Gillespie2018a,Jhaver2018a}, has become a crucial task.
Formally, we refer to videos that violate platform policies or guidelines as deviant videos, and crowd workers are recruited to review videos as moderators.

Manually reviewing the sheer volume of videos is a labor-intensive task for human moderators; thus, several mixed-initiative methods are employed to guard against deviant content~\cite{Hassner2012a}.
Such methods integrate human intelligence with computational power through a ``filter first and review later'' strategy, in which moderators review only suspicious videos recalled by machine learning models~\cite{Nievas2011a} (Fig.~\ref{fig:mixed-initiative-framework}).
Although many advanced machine learning models~\cite{He2016a,Mnih2015a,Wu2021survey} can achieve comparable or better performance than people in mixed-initiative systems, they may disrupt human moderators owing to their low accuracy for real-world videos and ``black box'' feature~\cite{Chollet2018a}.
For example, such models may generate uncertain or even erroneous results contradicting human judgement, thereby increasing the burden of moderators, who need to not only review the videos but also consider ``who'' is right.

To better leverage human moderators with computational power, we argue that reviewing multimodal video content while being aware of machine-generated risk information through interactive visualizations would be beneficial.
\trevise{Formally, we refer to risk as the possibility of deviant video content that disobey moderation policies.}
To validate our assumption, we collaborated with a leading domestic moderation company that employs thousands of moderators for daily video-reviewing tasks.
We first developed an alpha prototype that integrates a timeline visualization to demonstrate temporal risks and deployed it as an internal tool for the employees.
We then conducted a preliminary study to observe the moderators who adopted our tool to complete their works.
By comparing their daily and historical performances, we found that their average time efficiency increased by $22.1\%$, while the missing rate decreased by $81\%$.
The promising results motivated us to propose a risk-aware framework that deeply integrates machine-generated insights into the reviewing process through visualizations.
However, the development of such a framework involves two major obstacles:

\textbf{Extraction of multimodal risk information.}
The multimodal semantic content of videos can be successfully obtained using deep learning-based models, which can lead to many successful applications~\cite{He2016a,Xie2018a,Poco2017a}.
Nevertheless, such models cannot be directly employed in e-commerce videos owing to the lack of domain-specific datasets and the diversity of moderation policies.
The extraction of semantic risk content from e-commerce livestreaming videos requires a novel multimodal model, which remains an unsolved challenge.

\textbf{Visualization of risk-aware multimodal video content.}
The massive multimodal video content requires considerable human efforts to review entire videos, which is time consuming.
Moreover, extracted risk information is typically dynamic and high-dimensional, which may produce heavy visual clutter when being visualized with informative video context.
Thus, the integrated visualization of multimodal video content and associated risks poses the second challenge, which demands novel visual summarization techniques.

In this work, we propose VideoModerator, a risk-aware framework, to facilitate multimodal video moderation by using a mixed-initiative method.
To address the first challenge, we employ state-of-the-art deep learning models~\cite{He2016a,Szegedy2016a,Ren2017a,Moritz2020a} to extract risk tags from potentially deviant frames and detect risk words from audio scripts.
To overcome the lack of training data, we use a ``learning with reviewing'' strategy, in which reviewed videos with ground-truth labels are used to train deep learning models iteratively.
For the second challenge, we propose a multimodal visualization interface consisting of three components:
a) A \textit{video view} involves a video player integrated with a segmented timeline presenting the high-risk periods that moderators should focus on.
b) A \textit{frame view} involves visually summarized video frames to provide a compact overview at the frame-, shot-, and scene levels.
We project the frames on the vertical axis and cluster them to construct different shots and scenes.
The horizontal axis is the timeline and circular glyphs are introduced to visualize the associated risk tags. 
c) An \textit{audio view} involves the automatic translation of audio clips into a list of sentences that may contain risk words.
We employ a histogram to indicate the frequencies of the risk words and adopt a storyline-based layout to demonstrate the audio context.
With the proposed interface (Fig.~\ref{fig:video-moderator}), we provide moderators a simple and efficient workflow.
A moderator starts his/her task by first clicking the \textit{load} button.
The video view enables the moderator to notice high-risk periods promptly, and the frame and audio view provide detailed risk information to help him/her make a quick decision.
The moderator completes the task by selecting a color label and clicking the \textit{submit} button.

The main contributions are as follows:
\begin{itemize}[nosep]
    \item We introduce a risk-aware framework, namely, VideoModerator, to facilitate the efficient moderation of e-commerce videos.
    \item We propose novel visual summarization techniques to visualize multimodal video content and the risk-aware information.
    \item We conduct experiments and a controlled user study to evaluate VideoModerator and report its usage through a case scenario.
\end{itemize}

%% file: sec2_related.tex
\section{Related Work}
In this section, we summarize the key techniques proposed for video moderation and relevant studies on video visualization.

\subsection{Video Moderation}
Video moderation means reviewing videos to protect viewers from deviant content violating explicit rules and community guidelines~\cite{Roberts2016a}.
One intuitive moderation strategy is to allow viewers to report video feeds with deviant content and protect themselves by using blocklists~\cite{Lampe2004a,Jhaver2018a}.
However, the effectiveness of this method relies on user engagement with systems that may disrupt the normal usage of video platforms~\cite{Kiesler2012a}.
An extreme situation would involve all users becoming moderators, trapped in the fight against deviant content, rather than enjoying Internet surfing.
Thus, online platforms should adopt moderation services to protect their users and brand value~\cite{Chancellor2017a,Gupta2018a}.
Despite previous investigations on online deviant content, video moderation has garnered renewed attention owing to the emergence of state-of-the-art machine learning methods~\cite{Giannakopoulos2010a,Hanson2019a,Peixoto2019a,Wu2020a}.

To identify violent content in videos, Hanson et al.~\cite{Hanson2019a} introduced a spatiotemporal encoder built on the bidirectional convolutional LSTM architecture that generates improved video representations by leveraging spatial features with long-range temporal information.
Nevertheless, this solution takes videos as a whole without considering underlying violent concepts.
Peixoto et al.~\cite{Peixoto2019a} proposed a novel methodology that employs two separate deep neural networks to learn subjective- and conceptual-level spatiotemporal information.
Moreover, the authors aggregated deep representations~\cite{Yuan2021survey} under each specific concept by training a binary classifier to moderate videos.
Considering the multimodal features of videos, Wu et al.~\cite{Wu2020a} proposed a neural network that contains holistic, localized, and score branches to determine deviant videos by considering video and audio signals at different levels.
Meanwhile, Giannakopoulos et al.~\cite{Giannakopoulos2010a} employed a ``fusion'' strategy combining audio and visual information to moderate video segments.

Despite the advancements in deep learning-based video moderation techniques~\cite{Hassner2012a,Nievas2011a,Singh2018a}, using solely machine intelligence is insufficient owing to the limitations of model generalization~\cite{Chollet2018a} and the changes in moderation guidelines~\cite{Gillespie2018a}.
For many online moderation applications~\cite{MS,LionBridgeAI}, first, machine learning models are employed to recall underlying deviant videos; then, human moderators are recruited to review the videos closely.
This two-step strategy intends to achieve an improved trade-off between preventing irregular content and protecting wrongfully identified videos.
However, this strategy ignores the risk-aware information extracted by machine learning models, which is important for discovering deviant content.
To promote an improved moderation paradigm, we propose a risk-aware framework that leverages human moderators with the deep analysis of machine-generated insights.
To the best of our knowledge, this study is the first to employ visual analytic approaches to moderate multimodal video content.

\begin{figure}[t]
  \includegraphics[width=0.48\textwidth]{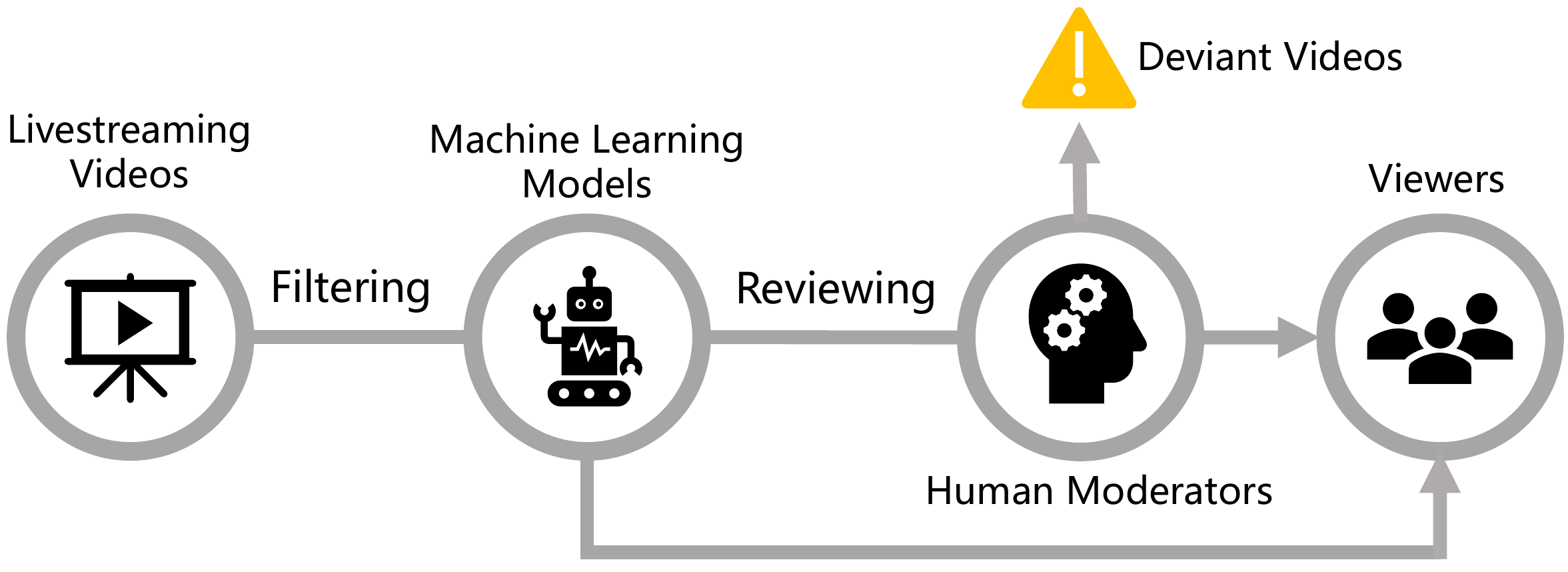}
  \caption{A typical mixed-initiative framework for video moderation.
  Videos are firstly filtered by machine learning models and the suspicious videos are further reviewed by human moderators.
  Normal videos would be presented to public directly while deviant videos would be handled by different community policies~\cite{Twitter, Youtube}.
  }
  \label{fig:mixed-initiative-framework}
  \vspace{-2mm}
\end{figure}

\subsection{Video Visualization}
Video visualization has been investigated in the fields of multimedia analysis~\cite{Chinchor2010a,Zahalka2014a,Jamonnak2021Geo} and information visualization~\cite{Borgo2012a,Daniel2003a,Soure2021CoUX,Lan2021RallyComparator,Wang2021Tac,Di2021Bus,Ye2021Shuttle,Wu2021Co,Xie2021PassVizor} for decades.
We consider the relevant studies on \textit{visual video summarization} and \textit{visual video analytics} that are associated with video reviewing.   

\subsubsection{Visual Video Summarization}
Visual video summarization refers to the process of transforming a raw video into a compact visual form without losing substantial information.
The goal of this process is to facilitate rapid video navigation by transforming video content and associated features into \textit{volume-}~\cite{Nguyen2012a,Kang2006a,Daniel2003a}, \textit{feature-}~\cite{Chen2006a,Duffy2015a,Tang2018a}, and \textit{context-}based visualizations~\cite{Romero2008a,Hoferlin2013a,Botchen2008a,Meghdadi2013a}.

Volume-based video visualizations, which are also known as \textit{video graphics}~\cite{Borgo2012a}, employ volume rendering to transform video frames into a single graphic.
Daniel et al.~\cite{Daniel2003a} formally proposed the idea of video visualization and employed different volume visualization techniques to visualize videos as volume cubes, which inspired several studies~\cite{Kang2006a,Nguyen2012a}.
For example, Kang et al.~\cite{Kang2006a} proposed a space-time video montage that represents a video by using volumetric layers and combines highly informative portions into an integrated visualization.
Nguyen et al.~\cite{Nguyen2012a} developed VideoSummagator, which transforms a video into a space-time cube by combining the most informative frames to facilitate a quick video overview.
Instead of visualizing raw videos, feature-based video visualizations focus on presenting extracted video features.
Chen et al.~\cite{Chen2006a} introduced visual signatures into video visualization and investigated different types of visual signatures through a user study.
Duffy et al.~\cite{Duffy2015a} proposed glyph-based video visualizations that employ tadpole-like glyphs to exhibit the spatiotemporal motion features of semen for semen analysis.
Storyline visualizations were also developed to understand the scenic interactions among characters for movie analysis~\cite{Tang2018a,Tang2020a}.
Context-based video visualizations have been established in feature-based visualizations with scene context.
Romero et al.~\cite{Romero2008a} introduced Viz-A-Vis, which visualizes human activities on a room map, and Botchen et al.~\cite{Botchen2008a} aggregated human motion trajectories on top of representative video frames.
For surveillance videos, H{\"o}ferlin et al.~\cite{Hoferlin2013a} developed schematic summaries aggregating trajectories on video scenes, and Meghdadi et al.~\cite{Meghdadi2013a} employed a space-time cube to layer the motion paths of moving objects on top of a video.

Despite these methods effectively summarize entire videos, they are established in various domains and designed for different purposes and cannot be directly applied in video moderation.
Moreover, they mainly focus on the visual aspect of videos while ignoring audio context.
This study focuses on visualizing multimodal video content and integrating risk-aware information to facilitate quick video navigation.

\subsubsection{Visual Video Analytics}
Visual video analytics refers to the reasoning process facilitated by a visualization interface to complete video analytic tasks~\cite{Chinchor2010a,Hoferlin2015a}.
Various visual analytic approaches have been developed for movies~\cite{Kurzhals2016a,Ma2020a}, news~\cite{John2019a,Renoust2017a}, presentations~\cite{Wu2020b,Zeng2020a}, and education~\cite{Zeng2020b,Guo2021DanceVis}.

Movies provide multimodal information that can assist analysts in understanding narratives.
To analyze such information, Kurzhals et al.~\cite{Kurzhals2016a} proposed multilayer segmented timelines to characterize the visual content of movies and associated video frames with semantic information derived from subtitles or scripts.
Meanwhile, Ma et al.~\cite{Ma2020a} focused on the emotional content of movies and employed a map metaphor for tracking and understanding the evolution of characters' emotions.
News videos are important for journalists to discover significant topics and emergent events.
John et al.~\cite{John2019a} utilized multiple coordinated views to associate news videos with textual reports, and Renoust et al.~\cite{Renoust2017a} proposed FaceCloud, which follows word clouds to present images and words in a compact view.
Presentation videos have become prevalent owing to their easy-to-access feature~\cite{Wu2020b,Zeng2020a}.
Wu et al.~\cite{Wu2020b} employed Sankey diagrams integrated with segmented timelines to demonstrate presentation skills adopted by speakers, and Zeng et al.~\cite{Zeng2020a} established a hybrid visualization combining speakers' faces and video captions to understand speakers' emotion coherence.
To assist educators in improving teaching quality, Zeng et al.~\cite{Zeng2020b} developed EmotionCues, which employs a novel flow visualization to track students' emotions and understand their mental states.

Although previous studies have utilized coordinated views to analyze multimodal video content, they are established in a typical analytic framework~\cite{Hoferlin2015a} that primarily focuses on the visualization side.
H{\"o}ferlin et al.~\cite{Hoferlin2012a} proposed an interactive learning framework, in which users can annotate videos in a mixed-initiative way.
Inspired by this framework, we propose a risk-aware framework that deeply integrates human knowledge with computational power by using novel interactive video visualizations to facilitate efficient video moderation.

%% file: sec3_framework.tex
\section{Background}
This section presents the background and data for video moderation in e-commerce platforms where vendors intend to advertise their products through online livestreaming~\cite{Cai2019a}.
On the one hand, such activities produce a sheer volume of videos that enable vendors to attract potential consumers and increase conversion rates.
On the other hand, user-generated videos may inevitably introduce potential risks due to the possibility of deviant content.
Thus, video moderation becomes an essential task for e-commerce platforms that employ numerous crowd workers, also known as moderators, to review and moderate videos.
We first describe the basic characteristics of e-commerce livestreamings and introduce e-commerce moderation policies.
We then provide a preliminary study which motivates us to complete this study.

\begin{figure}[t]
  \includegraphics[width=0.48\textwidth]{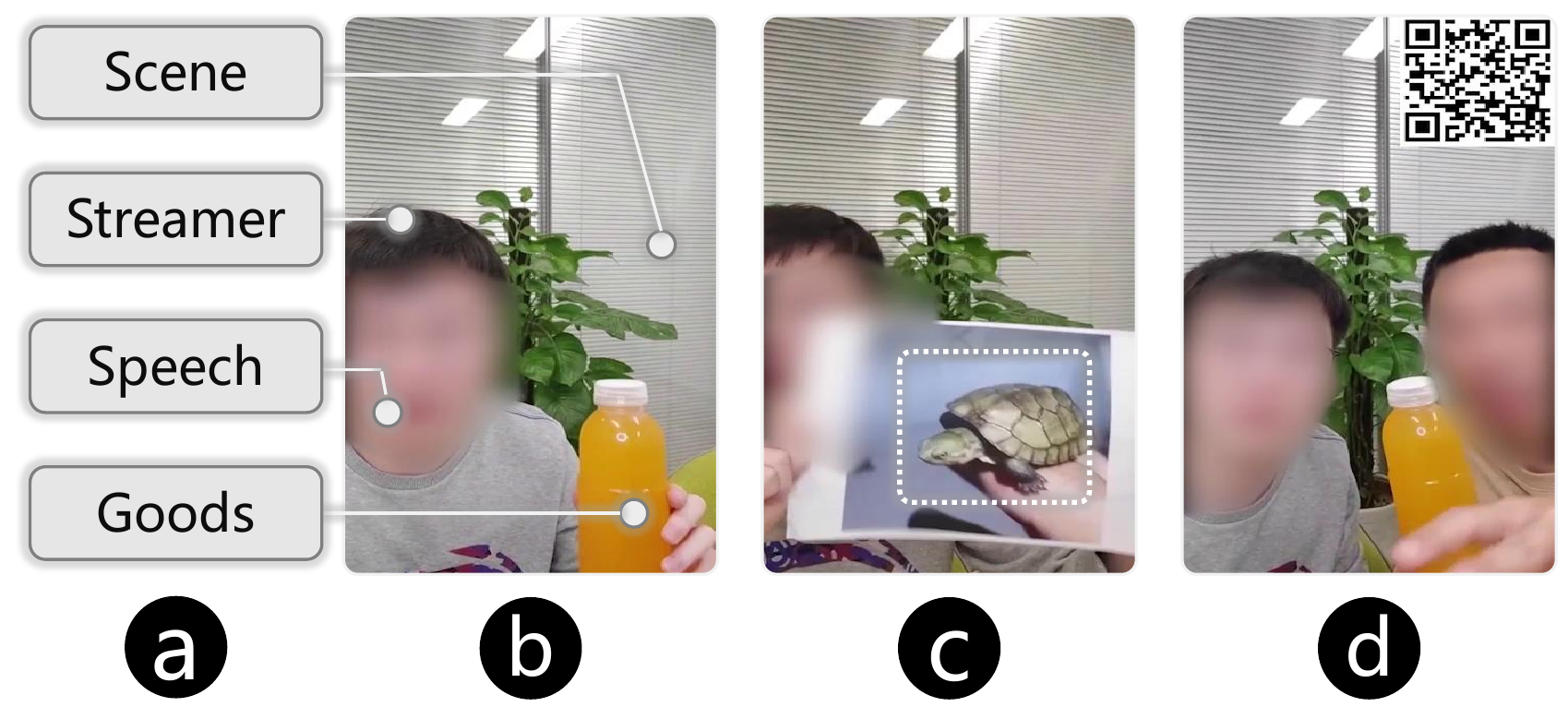}
  \caption{E-commerce livestreaming examples.
  (a) The four essential narrative elements of e-commerce livestreams.
  (b), (c), and (d) are frames violating the policies of false advertising, protected products, and inappropriate business.
  The dashed white box is manually annotated.
  }
  \label{fig:background}
  \vspace{-4mm}
\end{figure}

E-commerce livestreaming refers to the broadcasting of live videos in real time on the Internet to promote and sell goods to the public, which is regarded as the future of business and marketing.
Despite the online feature of livestreams, we focus on an off-line scenario in which moderators review ended livestreams to determine whether the livestreamers followed or violated platform policies.
Given that we focus only on ended livestreams, we do not distinguish between online and off-line videos in the rest of the paper.
E-commerce livestreaming videos can be characterized by four narrative elements, namely, the \textit{streamer}, \textit{goods}, \textit{scene}, and \textit{speech} (Fig.~\ref{fig:background}).
\textit{Streamer} refers to the speaker who is the main ``actor'' in a livestream.
Occasionally, two or more streamers, who are often small business owners on e-commerce platforms, may appear in a single livestream.
\textit{Goods} represent the products promoted by a streamer, and \textit{scene} is the location of a livestream, which is typically recorded against a fixed background.
Lastly, \textit{speech} refers to what a streamer is saying to advertise his/her goods.
E-commerce livestreaming videos are typical multimodal data.

Generally, e-commerce videos can be divided into two types, namely, \textit{normal} and \textit{deviant}. Normal videos are presented directly to viewers, whereas deviant videos are handled by different community guidelines or policies.
For example, YouTube may take down copyright-violated videos~\cite{Youtube}, and Twitter may suspend accounts that upload videos containing violence or threats~\cite{Twitter}.
To understand moderation policies for e-commerce livestreaming videos fully, we collaborate with two domain experts who have been working in a leading e-commerce company for decades.
Four common moderation policies, namely, \textit{false advertising}, \textit{protected products}, \textit{inappropriate business}, and \textit{sensitive content}, exist; they are designed to protect platforms' business and brand value.
The relationships between the moderation policies and narrative elements are explained below:
\begin{itemize}[nosep]
  \item \ttfa{False advertising}
  refers to the usage of exaggerated statements or false claims on properties, goods, or services for the public.
  Moderators typically pay attention to the \textit{goods} and \textit{speech} elements to examine whether streamers use misleading words.
  \item \ttpp{Protected products}
  are goods protected by laws, such as smuggled goods, protected animals or plants.
  Moderators mainly focus on the \textit{goods} element to determine whether an advertised product belongs to the white or black list.
  \item \ttib{Inappropriate business}
  refers to the deviant behaviors of streamers who deliberately ruin platform publicity.
  Moderators usually examine the \textit{scene} element to detect whether pictures or slogans aim to ruin brand image deliberately.
  \item \ttsc{Sensitive content}
  refers to violent, hateful, or explicit content inappropriate for all-ages e-commerce platforms.
  Moderators should pay attention to the \textit{streamer}, \textit{scene}, and \textit{speech} elements to determine whether deviant content exists.
\end{itemize}

Owing to the diversity of videos, reviewing videos by relying solely on high-level policies is difficult for moderators.
According to experts, we define dozens of specific risk tags and words for each policy to moderate videos and audio clips, respectively.
We refer to each policy as a risk category and obtain more than $100$ risk tags and words for 4 risk categories.
To moderate a video, moderators must first review the entire video to check for deviant content and then label the risk category to indicate which policy it violates.
Moreover, moderators must submit evidence supporting their conclusion, including at least one screenshot of the deviant content and associated risk tags or words, to help streamers improve their livestreams.

\section{Risk-aware Framework}
This section first presents a preliminary study that motivates our risk-aware framework for multimodal video moderation.
A detailed description of the framework and its two back-end components, namely, the data-processing procedure and data-filtering method, is followed.

\begin{figure*}[t]
  \includegraphics[width=0.98\textwidth]{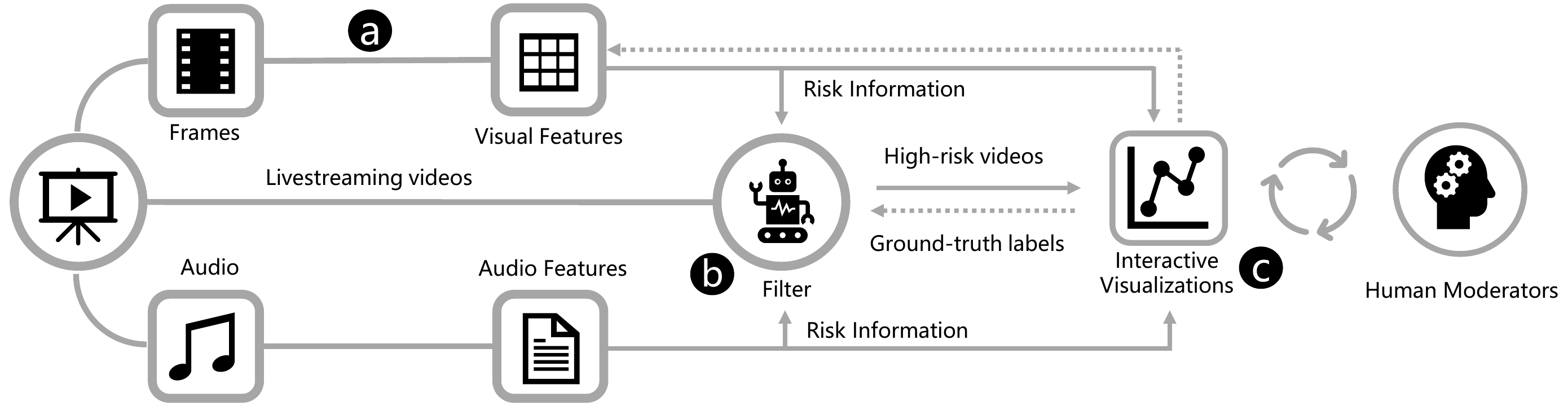}
  \caption{The risk-aware framework for video moderation:
  (a) \textit{video processing} employs multimodal techniques to process the audio and visual information of livestreaming videos separately.
  (b) \textit{video filtering} intends to select possibly deviant videos which would be further reviewed by moderators.
  (c) \textit{video reviewing} employs interactive visualizations to facilitate efficient video moderation by keeping moderators aware of risks.
  The solid lines refer to video data and associated risk information, whereas the dashed lines refer to ground-truth labels provided by moderators.
  }
  \label{fig:risk-aware-framework}
\end{figure*}

\subsection{Preliminary Study}
The existing video moderation framework (Fig.~\ref{fig:mixed-initiative-framework}) integrates a generic reviewing tool~\cite{MS} consisting of a video player with small multiples of video frames (Fig.~\ref{fig:baseline}).
The basic idea behind this framework is that reviewing less information means better time performance.
Thus, moderators are required to review original video content with minimum additional information, such as machine-generated scores.
However, we present a different assumption that enabling moderators to explore multimodal video content and extracted risk information concurrently may also improve their performance.
We conducted a preliminary study in collaboration with a moderation group in an e-commerce company to validate this assumption.

\begin{figure*}[t]
  \includegraphics[width=\textwidth]{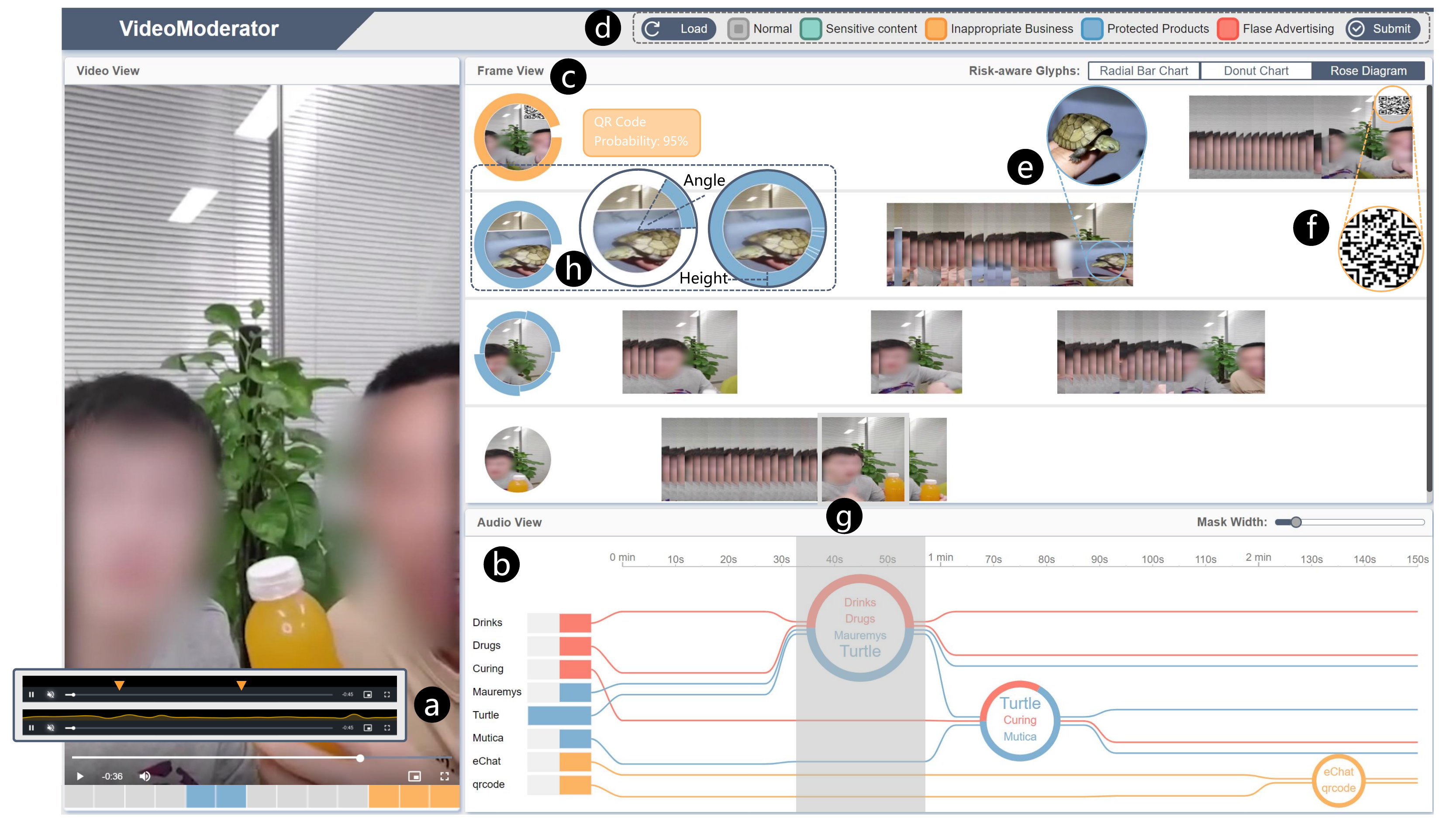}
  \caption{
    Multimodal user interface of VideoModerator:
    (a) a video view with a segmented timeline that demonstrates risk distributions;
    (b) an audio view with a combination of histograms and a storyline-based words visualization;
    (c) a frame view that summarizes the video frames;
    (d) a control panel that integrates a color legend encoding four risk categories;
    (e) and (f) are discovered insights for video moderation;
    (g) a moving window that visually associates audio and frame content;
    (h) circular glyphs that visualize risk-aware information. 
    The dashed lines are manually added.
  }
  \label{fig:video-moderator}
  \vspace{-5mm}
\end{figure*}

\trevise{
\textbf{Apparatus.}
We adopted the moderation tool that was used in the e-commerce company as the baseline interface, which was similar to Fig.~\ref{fig:baseline}.
To illustrate the benefits of integrating machine-generated insights, we enhanced the baseline interface by providing a segmented risk timeline (Fig.~\ref{fig:video-moderator}a).
Specifically, we divided the timeline into different segments and visualized their risk categories using the scores extracted by machine learning models.
Detailed discussions can be found in Sec. 5.2.1.
Except for the risk timeline, both interfaces shared the same components, namely, a video player and a photo wall presenting key frames.
We deployed the enhanced interface on the internal moderation platform of the e-commerce company.
}

\trevise{
\textbf{Participants and Procedure.}
We invited $80$ professional moderators to complete their daily tasks on the enhanced interface and collected their working logs for 1 week ($5$ days).
Generally, an e-commerce livestreaming lasts several hours that can be divided into hundreds of video clips and each moderator would review more than $10,000$ video clips every day.
Owing to the large number of reviewed videos, we sampled $10\%$ videos to evaluate the performance of moderators.
The sampled videos would be reviewed again and randomly assigned to the moderators.
The final reviews were regarded as ground truth.
}

\trevise{
\textbf{Metrics and Results.}
We employed two crucial metrics, namely, time efficiency (\textbf{TE}) and missing rate (\textbf{MR}).
Time efficiency refers to the number of videos that a moderator can review during a certain time.
Missing rate indicates the ratio of deviant videos that a moderator cannot discover in a predefined group of videos.
They are two primary metrics because a higher TE indicates the expansion of their business, while a lower MR ensures the strong control of risk level.
We obtained the two metrics by analyzing the collected working logs and checking the duplicate reviews of the sampled videos. 
Compared with the historical statistics provided by our collaborators, we found that our risk-aware visualization could improve the average time efficiency of the moderators by $22.1\%$ and decrease their average missing rate by $81\%$. 
Despite this pilot study is established on the video view, it motivates us to develop more visualizations that integrate human knowledge and machine intelligence deeply.
}

\subsection{Framework}
To handle the sheer volume of videos, existing systems~\cite{MS,LionBridgeAI} follow a typical framework that empowers human moderators with computational models (Fig.~\ref{fig:mixed-initiative-framework}).
Machine learning models were initially employed to detect potentially deviant video content violating community guidelines or policies.
Once deviant clips were detected, an entire video would be recalled for further review, whereas normal videos would be presented to the public. 
Although machine learning models can drastically reduce the number of videos to be moderated, two bottlenecks exist in the typical mixed-initiative framework.
First, the human reviewing process is hindered by the linear reading habit.
Moderators browse videos linearly to avoid skipping deviant content, which is labor intensive and time consuming.
Second, the complexity and diversity of real-world video data can decrease model accuracy.
Moderators need to review a large amount of normal videos wrongly selected by models, which increases their reviewing efforts.

To overcome these obstacles, we propose a risk-aware framework that deeply integrates human knowledge with machine-generated insights through interactive visualizations~\cite{Andrienko2021,Han2021NetV}.
As shown in Fig.~\ref{fig:risk-aware-framework}, our framework involves three steps, namely, \textit{video processing}, \textit{video filtering}, and \textit{video reviewing}.
The novelty of our framework lies in two aspects.
First, we consider the multimodal video content and risk information extracted using machine learning models.
Owing to the emergence of deep learning techniques~\cite{He2016a,Ren2017a}, we employ state-of-the-art models\cite{He2016a,Szegedy2016a,Ren2017a,Moritz2020a} to detect objects in video frames and translate speech from audio content.
Such information contains uncertainties, such as risk tags and the associated probabilities, which are usually ignored by the typical framework.
Our framework not only visualizes the multimodal video content but also presents such risk information to ensure effective video moderation.
Second, we establish a tight connection between machine learning models and human moderators by providing easy-to-use interactions. 
To reduce the number of videos to be moderated, we train a binary classifier to select high-risk videos that may include deviant content automatically.
The typical framework employs a ``filtering first and reviewing later'' method, in which pretrained models are adopted as classifier, whose accuracy may decrease considerably for new videos.
Our framework employs a ``learning with reviewing'' strategy, in which newly reviewed videos are used as ground-truth to update the training process of the classifier periodically.

\subsection{Video Processing}
The data-processing procedure comprises video-audio decomposition and feature extraction.
Our framework first separates a video into a sequence of frames and audio clips and then employs state-of-the-art techniques~\cite{He2016a,Szegedy2016a,Ren2017a,Moritz2020a} to extract multimodal features.

\textbf{Visual Feature Extraction.}
We extract visual features at frame and video levels using deep learning-based techniques~\cite{He2016a,Szegedy2016a,Ren2017a}.
\trevise{
An input video would be first decomposed into a sequence of video frames that contain redundant information.
We employ an adaptive sampling strategy~\cite{Zhao2021a} to reduce the time cost of the processing procedure.
Specifically, we sample one frame every 1 s for short videos (e.g., less than $30$ min) and one frame every 2 s for long videos.
We employ object detection~\cite{Ren2017a} and classification~\cite{He2016a} methods to generate the risk tags and scores for the sampled frames.
The first model~\cite{Ren2017a} employs a region proposal network to predict object bounds and scores and the second model~\cite{He2016a} proposes a deep residual neural network to facilitate image recognition task.
However, there are some risks that could not be detected using the frame-level information, such as the deviant behaviors of streamers.
Thus, we also divide a video into a set of video clips and adopt the inception model~\cite{Szegedy2016a} to obtain the video-level risk tags.
Given that there are numerous inherently different risk tags, we divide them into different groups and train every model to separately predict the risk tags within a group.
For example, the object detection model~\cite{Ren2017a} is trained to only generate the risk tags about protected products.
We employ an aggregation method to combine the risk tags generated by different models.
Each video $V$ is represented by a sequence of frames $\{F_1, F_2, ..., F_n\}$, and each frame is depicted by a set of risk tags $\{a_1, a_2, ..., a_s\}$.
We denote a frame as $F_n=(p_1, p_2, ..., p_s)$, where $p_s=\sum_i p_s^i$ and $p_s^i$ refer to the predicted score of risk tag $a_s$ by the $i$-th model.
Especially, $p_s^i=0$ if the risk tag $a_s$ cannot be detected by the model $i$.
}
The training of these advanced models requires a large amount of e-commerce livestreaming videos with ground-truth labels, which cannot be obtained from existing public datasets.
We first adopt pretrained models to initiate the framework and then employ a ``learning with reviewing'' method, in which the reviewed videos are collected to re-train the models.

\textbf{Audio Feature Extraction.}
We extract audio features at different regular time intervals to preserve the time context of streamers' speech.
Considering moderators' linear reading habit in receiving audio information, we employ the speech-to-text technique to support the visual exploration of streamers' words.
Specifically, we first adopt the automatic speech recognition (ASR) method~\cite{Moritz2020a} to translate spoken language into text and then detect potentially deviant content by using a list of risk words.
\trevise{
This method~\cite{Moritz2020a} proposes a transformer based streaming ASR system to generate output shortly after each spoken word, which can be applied to video applications.
}
To obtain sufficient training data, we automatically extract audio clips and detect textual captions from movies.
Such audio and text pairs can be used as ground-truth datasets to train the advanced ASR model~\cite{Moritz2020a}.
\trevise{
We extract audio features by calculating the frequencies of risk words detected from the input video. 
}
Each video $V$ is divided into a set of audio clips $\{A_1, A_2, ..., A_m\}$ at regular intervals.
\trevise{With the ASR model, each audio clip $A_m$ is translated into a bag of words, in which risk words $\{w_1, w_2, ..., w_t\}$ are identified.
We denote an audio clip as vector $A_m=(v_1, v_2, ..., v_t)$ where $v_t$ refers to the frequency of word $w_t$.
}

\subsection{Video Filtering}
To moderate the sheer volume of videos, we adopt a binary classifier as the filter, which can discover potentially deviant videos in accordance with extracted visual and audio features.
However, video moderation policies differ across platforms and applications.
Obtaining sufficient data to train a classifier to classify e-commerce livestreaming videos is difficult.
Following the idea of ``learning with reviewing,'' we first define a linear classifier to initiate the framework and then adopt the reviewed videos as the ground truths to improve the classifier iteratively.

Given that extracted features demonstrate how videos may potentially contain deviant clips, we define the risk value $R$ for each video $V$ by accumulating the probabilistic information of risk tags and words.
\trevise{We notice that the visual and audio features are extracted by different models.
To aggregate them, we first normalize the predicted scores that $p_s, v_t\in[0,1]$ then employ a linear combination method as following:
}
\begin{equation}
  R(V)=\frac{1}{m + n} (\sum_{i=1}^{n} R_F(F_i) + \sum_{j=1}^{m} R_A(A_j))
  \label{eq:risk-value}
\end{equation}

where $R_F$ and $R_A$ refer to the risk values of video frames and audio clips, respectively.
An intuitive strategy is to adopt the average values of $F_i$ and $A_j$ to define $R_F$ and $R_A$.
Nonetheless, the vectors of risk tags or words are often sparse, which may underestimate the risk values of possible deviant videos.
Thus, we employ a \textit{max aggregation} strategy that defines $R_F(F_i) = \max_s F_i^s$ and $R_A(A_j) = \max_t A_j^t$ where $F_i^s$ and $A_j^t$ refer to the $s$-th value of the $i$-th frame and $t$-th vector value of the $j$-th audio vector, respectively.
In accordance with Eq.~\ref{eq:risk-value}, we refer to the videos with risk values above the threshold $\bar{R}$ as high-risk videos.

The linear classifier may be inadequate for classifying diverse and complex video data. 
After obtaining enough reviewed videos, we adopt a state-of-the-art neural network~\cite{He2016a} as the filter, which can fit a nonlinear hyperplane.
Instead of obtaining the original video content, our classifier regards the extracted features as the input and is then trained using the moderator-reviewed labels.
\trevise{
The input data can be formulated as two matrices $(F_1,F_2,...,F_n)$ and $(A_1,A_2,...,A_n)$ that represent visual and audio features, respectively.  
}

%% file: sec4_visualdesign.tex
\section{Visualization Techniques}
This section describes \trevise{the design process} and considerations for VideoModerator followed by the visual designs and interactions.

\subsection{Design Process and Considerations}
\trevise{
To conduct the pilot study, we first developed the video view that illustrated the benefits of integrating human knowledge and machine intelligence.
After that, we further took a step to work closely with two domain experts.
They were also involved in the discussion of moderation policies and designed the baseline moderation tool that was used in the company.
Through regular meetings and discussions, we derived two high-level design considerations (DCs) followed by a set of specific design requirements (R).
}

\noindent
\textbf{DC1} \textbf{Visual Summarization of Multimodal Video Content.}

\begin{enumerate}[nosep]
  \renewcommand{\labelenumi}{\textbf{R\theenumi}}
  \item \textit{Highlighting high-risk periods in a video.}
  Browsing a video stream without visual hints may be tedious.
  Thus, the system should automatically detect and highlight critical periods that moderators should consider in priority order.
  \item \textit{Presenting a multilevel overview of frames.}
  Long videos contain numerous frames that can hinder the exploration of video content.
  Hence, the system should provide a scalable and compact summarization of frames to support an effective overview.
  \item \textit{Displaying a multifaceted overview of audio content.}
  The linear reading habit makes searching audio clips time consuming.
  Therefore, the system should provide a multifaceted overview to enable the efficient exploration of audio content.
\end{enumerate}

\noindent
\textbf{DC2} \textbf{In-depth Exploration of Risk Information.}

\begin{enumerate}[nosep]
  \setcounter{enumi}{3}
  \renewcommand{\labelenumi}{\textbf{R\theenumi}}
  \item \textit{Contextualizing risk information with multimodal content.}
  Risk information is extracted from video or audio content.
  The system should combine multimodal content with associated risks to provide concrete visual representations.
  \item \textit{Linking associated risk-aware visualizations in coordinated views.}
  A gap exists in the risk information extracted from multimodal content.
  The system should visually connect risk-aware visualizations in different views to facilitate quick labeling.
\end{enumerate}

\subsection{Visual Design}
Figure~\ref{fig:video-moderator} shows our interface with three major views, that is,
a video view for navigating videos and identifying high-risk periods (Fig.~\ref{fig:video-moderator}a for \textbf{R1}),
an audio view for displaying a multifaceted risk-aware audio overview (Fig.~\ref{fig:video-moderator}b for \textbf{R2} and \textbf{R4}),
and a frame view for providing a compact visual representation of videos (Fig.~\ref{fig:video-moderator}c for \textbf{R3} and \textbf{R4}).
A set of easy-to-use interactions is developed to link the associated risk-aware visualizations in coordinated views (\textbf{R5}).

\subsubsection{Video View}
The video view consists of a video player and a segmented timeline that displays the risk distribution of different time periods.
The video player integrates a set of common video tools (e.g., \textit{play/stop} and \textit{fast-forward}) to assist the moderators in flexibly browsing the video content.
To \textit{highlight high-risk periods in a video} (\textbf{R1}), we propose three alternative designs (Fig.~\ref{fig:video-moderator}a from the bottom to the top), namely, a segmented timeline with risk indicators, a timeline with in-line flow visualization, and a timeline with triangle glyphs.
The design with triangular glyphs merely highlights the high-risk moments but lacks the necessary risk context, which may lead to the skipping of potentially deviant frames.
To overcome this limitation, we insert a flow chart depicting the risk distribution of an entire video into the timeline.
Nevertheless, the flow chart is excessively informative to highlight high-risk moments effectively.
To balance the two extremes, we adopt a segmented timeline that is divided into different periods, in which the associated risk categories are indicated by color blocks.
We encode the risk categories corresponding to the four moderation policies by using a color scheme extracted from ColorBrewer~\cite{ColorBrewer} (see Fig.~\ref{fig:video-moderator}d).


\subsubsection{Frame View}
\begin{figure}[t]
  \centering
  \includegraphics[width=0.46\textwidth]{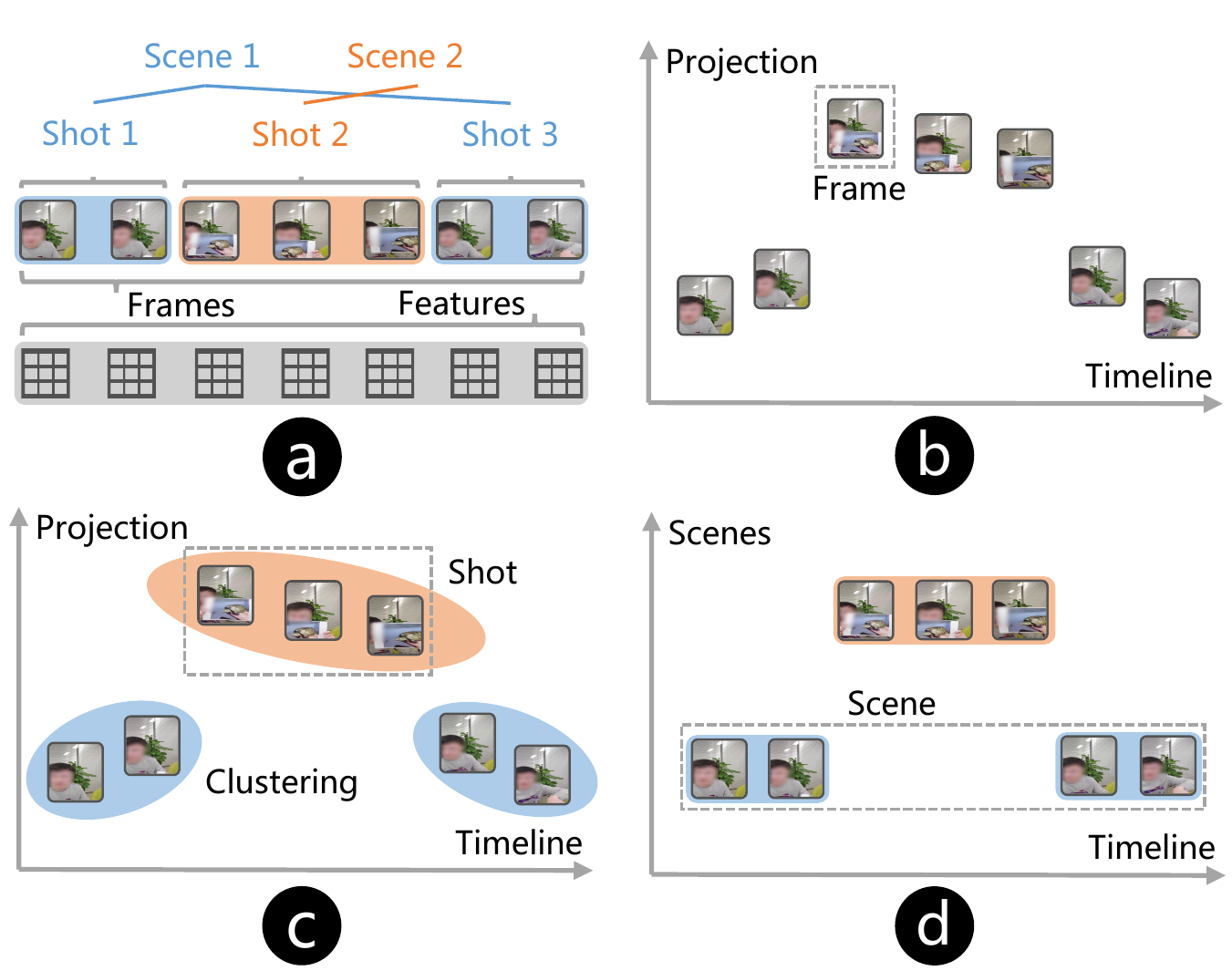}
  \caption{The visual video summarization method:
  (a) \trevise{input:} video frames and the extracted risk information;
  (b) \trevise{frame projection}: video frames are projected on the vertical axis using tSNE;
  (c) \trevise{shot clustering}: the projected frames are clustered using the density-based clustering technique;
  (d) \trevise{scene alignment: the similar shots are aligned to construct different scenes}.
  The vertical axis represents the 1D projection space in (b), (c) or scenes in (d).
  The horizontal axis refers to the timeline.
  }
  \label{fig:frame-layout}
  \vspace{-4mm}
\end{figure}

The frame comprises a multilevel overview of the video frames (\textbf{R2}) and a risk-aware circular glyph (\textbf{R4}) (see Fig.~\ref{fig:video-moderator}c).
To provide a concrete video summarization, we employ a well-established narrative model~\cite{Kurzhals2016a,Ma2020a} that characterizes video content at \textit{scene}, \textit{shot}, and \textit{frame} levels.
\trevise{
A shot refers to a set of similar consecutive frames (the orange box in Fig.~\ref{fig:frame-layout}a), and a scene refers to a list of similar shots (the two blue boxes in Fig.~\ref{fig:frame-layout}a).
Moreover, we adopt circular glyphs that surrounds a video thumbnail with risk-aware visualizations to keep moderators aware of the risks (Fig.~\ref{fig:video-moderator}h).
}

A video can be regarded as a linear visual signal consisting of consecutive frames.
To accelerate video browsing, representative video frames have been extracted to present most informative content by using a list~\cite{Haubold2005a} or small multiples~\cite{Hutchison2013a}, which is known as \textit{video summarization}~\cite{Ying2006a}.
However, these approaches may not be applicable to video moderation due to the gap between informative content and associated risks.
For example, the frame with the highest risk value may not be regarded as informative by state-of-the-art methods~\cite{Hutchison2013a,Gygli2014a}.
To avoid skipping potentially deviant frames, we propose a novel visual summarization method that organizes video frames in an intuitive and compact layout.
Our goal is to convert the reviewing of redundant video content into the browsing of several groups of video frames, which can be completed in a short period of time.
Specifically, we employ a projection method to align video frames along the temporal dimension and adopt a clustering technique to group the projected frames visually.

Our approach regards the sampled frames and extracted risk information as the input and processes these data by using three steps:
\trevise{a) \textit{frame projection}:} video frames are typical high-dimensional data that can be projected into a low-dimensional space in accordance with their spatial similarity~\cite{Van2008a} or semantic proximity~\cite{Xie2018a}.
An intuitive way to visualize video frames is to project them into a 2D space that can facilitate the visual grouping of the frames. Nonetheless, 2D-based projections ignore the temporal context of video frames, which is important for discovering shot-level insights.
Thus, we adopt the idea of temporal MDS plot~\cite{Jackle2016a}, which can project the frames on the vertical axis.
\trevise{The horizontal axis represents the temporal dimension.}
Instead of using the MDS method~\cite{Cox2008a}, we employ the tSNE technique~\cite{Van2008a} to project the video frames (Fig.~\ref{fig:frame-layout}b) due to its strength in creating tight frame groups for visualization.
\trevise{b) \textit{shot clustering}:} once obtaining the projected frames, we detect the shots consisting of a set of similar frames.
Considering that the relative vertical differences among the projected frames indicate their similarity, we further cluster them in accordance with their vertical positions (Fig.~\ref{fig:frame-layout}c).
Specifically, we employ the density-based clustering technique to group the video frames heuristically, in which each shot refers to a cluster of consecutive frames.
\trevise{c) \textit{scene alignment}: we further group the shots to construct scenes according to their vertical positions.
As shown in Fig.~\ref{fig:frame-layout}d, we align the shots whose relative vertical distances are less than a threshold and a scene is constructed by a set of shots on the same horizontal level.
}


\trevise{
Given a scene, we further visualize and aggregate the frame risks to assist moderators in discovering deviant content promptly.
Each scene $\overline{S}$ contains multiple frames $\{F_i\}_{i \in {\overline{S}}}$.
The risks of the $i$-th frame can be defined using vector $F_i=(p_1,p_2,...,p_s)$, where $p_s$ indicates the score of risk tag $a_s$.
We aggregate the frame-based risks within a scene by $R(\overline{S})=\sum_{i \in {\overline{S}}}F_i$.
As shown in Fig.~\ref{fig:video-moderator}h, we consider three alternative designs, namely, rose diagram, radial bar chart, and donut chart, to visualize $R(\overline{S})$.
Moreover, we integrate the circular designs~\cite{Filipov2021Circles} with a representative frame whose risk score is the highest in the scene (\textbf{R4}).
The radial bar chart visualizes risk tags by using bar sectors and the bar height to encode the risk score, which is easy to interpret.
The donut chart encodes the risk score by using the sector angle, which is scalable to sparse risk information.
The rose diagram encodes the risk score with the sector area, which highlights extreme risk scores effectively.
Considering that the three designs demonstrate their strengths in different tasks, we provide a radio button to enable users to select different designs in accordance with their requirements.
}

\subsubsection{Audio View}
The audio view provides a multifaceted overview of streamers' speech (\textbf{R3}), which consists of a horizontal histogram and a storyline-based visualization (Fig.~\ref{fig:video-moderator}b).
The horizontal histogram is designed to demonstrate the frequency distribution of the risk words extracted from the audio content, in which each bar refers to a word.
The height of a bar represents the count of the word mentioned in the speech, and its color represents the risk category, which is the same as the circular glyph.
We rank the risk words in descending order to enable moderators to review the risky content from highest to lowest.
However, moderating audio content by only reviewing the risk word summaries is insufficient, because moderators cannot make a reliable decision without the temporal context.
Thus, we adopt storyline-based visualization to reveal the temporal concurrence of risk words.

\begin{figure}[t]
  \centering
  \includegraphics[width=0.48\textwidth]{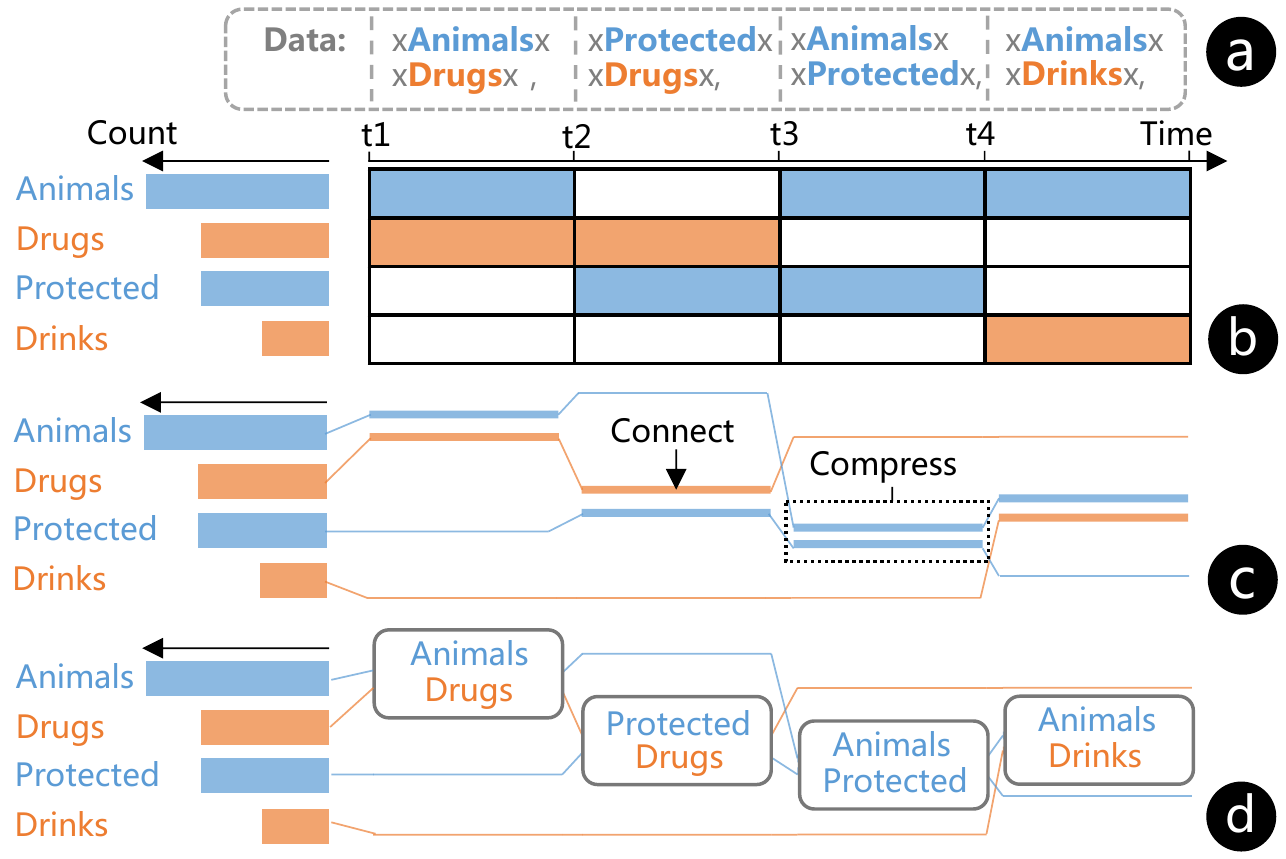}
  \caption{The visual summarization of audio content:
  (a) the input data contains four risk words;
  (b) a histogram and a segmented timeline to demonstrate \textit{when} and \textit{which} risk word is spoken;
  (c) a storyline layout is employed to connect line segments and compress redundant vertical space;
  (d) word clouds are adopted to provide specific context.
  }
  \label{fig:audio-view}
  \vspace{-4mm}
\end{figure}

Audio content can be regarded as a sequence of sentences translated using an ASR technique~\cite{Moritz2020a}.
Each sentence can be denoted as a tuple $(start\_time, end\_time, words)$ (Fig.~\ref{fig:audio-view}a), in which some \textit{words} are considered risky if they violate moderation policies.
To review audio content, moderators must know not only \textit{which} risk word but also \textit{when} it is mentioned.
They must also know \textit{what} other risk words are spoken concurrently.
We consider several alternative designs to provide such a complex context.
We consider a segmented timeline, in which each line represents a risk word and each filled segment indicates a word spoken during a certain period (Fig.~\ref{fig:audio-view}b).
The strength of the segmented timeline lies in its ability to demonstrate \textit{which} risk word and \textit{when} it is spoken effectively.
Understanding concurrent words in the same period is easy, but tracking the historical context may be difficult owing to the line segment disconnectivity.
Moreover, the design may cost redundant vertical space when risk words are sparse.
To overcome these limitations, we first compress the vertical space of the line segments then connect the segments of the same words to make them trackable (Fig.~\ref{fig:audio-view}c).
The connection of line segments may produce heavy visual clutter, as it increases line crossings and wiggles.
To obtain a legible and aesthetic layout, we employ the storyline algorithm to minimize the line crossings and wiggles~\cite{Tang2018a,Tang2020a}.
Moreover, we visualize the risk words using a word cloud to replace line groups to enable moderators to obtain a highly intuitive and comprehensive context (Fig.~\ref{fig:audio-view}d).

\subsection{Interactions}
In addition to basic interactions, such as \textit{playing or fast-forwarding a video}, VideoModerator integrates several other interactions to facilitate the fast reviewing of multimodal video content (\textbf{R5}).

\begin{itemize}[nosep]
  \item \textbf{Locating high-risk periods.}
  Moderators can locate high-risk periods by using the segmented timeline in Fig.~\ref{fig:video-moderator}a. 
  The system will update the video content during the located period when moderators click on the blocks.
  \item \textbf{Linking frames with risk tags.}
  Moderators can link video frames with associated risk tags to review deviant content quickly.
  In Fig.~\ref{fig:video-moderator}b, the system will automatically enlarge the associated frames when moderators hover over the sectors of the risk glyph.
  Moreover, moderators can update video content by clicking on a frame to locate deviant content promptly.
  \item \textbf{Exploring audio context.}
  Moderators can effectively explore audio context by clicking on a risk word or a histogram bar in Fig.~\ref{fig:video-moderator}c.
  The system will enlarge the links to highlight the associated temporal context.
  Moreover, moderators can click on a word cloud to specify the period they want to review further.
\end{itemize}

%% file: sec5_cases.tex
\section{Usage Scenario}
This section presents a usage scenario to demonstrate the usability~\cite{Wang2021Perception} and usefulness of VideoModerator.
We describe how Alan, an experienced moderator, reviews multimodal video content to discover deviant clips.
He starts his task by loading an e-commerce video, in which two streamers are advertising their orange juice (Fig.~\ref{fig:video-moderator}).
After loading the video, he notices that the segmented timeline immediately demonstrates the risk distribution across the entire video, whose deviant clips may be in the middle part and second half (Fig.~\ref{fig:video-moderator}a).
He then turns to the frame view and observes that the video is divided into four scenes, namely, the first scene with the two streamers and a QR code, the second scene with a turtle picture, the third scene with a tree background, and the fourth scene with a drink (from top to bottom).
He first examines the top scene because it shows the highest risk (Fig.~\ref{fig:video-moderator}f).
He immediately finds that the QR code disobeys the policy of inappropriate business because it links to a competitive e-commerce platform.
Once confirming one deviant clip and associated policy, he could choose to complete the task by clicking the label button (Fig.~\ref{fig:video-moderator}d).

However, he is interested in whether this video violates other polices and decides to examine the second scene wherein a turtle picture is shown.
He has no clues about the potential risk of the shown picture, so he hovers the rose diagram and finds that this is a protected turtle in accordance with the shown tooltip (Fig.~\ref{fig:video-moderator}e).
Thus, he confirms that the video also disobeys the policy of protected animals that are not allowed to sell.
He also discovers a possible deviant clip that shows no risk in the frame view (Fig.~\ref{fig:video-moderator}g).
Therefore, he turns to the audio view and examines the most left word cloud.
He notices that the streamers claim that their product can cure any disease, which is beyond the scope of a drink and violates the policy of false advertising.
Owing to the benefits of multimodal visualizations, he has discovered all deviant clips and determined to finish the task by assigning the most apparent risk label.
The online demo can be found at \textcolor{tblue}{\url{https://videomoderator.github.io/}}.

\section{Experiments}
\textbf{Implementation}
We employ a client-server architecture to implement our mixed-initiative framework. A back-end server is developed to integrate various machine learning models, and a visualization interface is developed to facilitate multimodal video moderation. We implement the interface by using React.js, and develop visualizations by using D3.js.
We implement the back-end server in Python and develop the machine learning models with a state-of-the-art framework, namely, PyTorch.
We deploy the back end on cloud computing devices and connect the client and server by using the web protocol HTTP.

\trevise{
We conducted online experiments to understand the performance of the machine learning models incorporated in VideoModerator.
The main goal of the experiments is to validate the effectiveness of the general framework instead of evaluating specific machine learning models.
By collaborating with the e-commerce company, we deployed our back-end on high-performance computing clusters powered by elastic algorithm services~\footnote{https://www.alibabacloud.com/}.
We trained our models with the livestreaming videos reviewed by moderators to detect sensitive content, including abusing (audio) and explicit visual content (video).
There were $21$ risk tags associated to explicit visual content and $11$ risk words related to abusing sentences.
We mainly focused on these specific risks because covering full risks required a large amount of computational resources which were not accessible in the experiments. 
}

\trevise{
The trained models were tested using the real-world videos on our collaborators' moderation platform.
In total, our models processed around $100$ thousands live streaming videos, including $0.4$ billions video frames and $36$ millions audio clips.
The average accuracy of the models are $33.2\%$ and $35\%$ for detecting audio and video risks, respectively.
The recalling rate of the binary classifier is $99.5\%$ for filtering possible deviant videos with abusing or explicit content.
The model accuracy is far from the expectations due to the diversity and complexity of real-world video data.
However, they are still acceptable according to our collaborators because platforms intend to have higher recalling rates, even by sacrificing model accuracy in real practices.
This strategy is understandable because the main goal of e-commerce platforms is to protect not prevent user-generated content.
Thus, moderators would like to have things under control and review ``possibly'' deviant videos to avoid wrongly-detect videos as much as possible.
}

%% file: sec6_userstudy.tex
\section{User Evaluation}
This section presents an in-lab user study to evaluate the user interface of VideoModerator by comparing it with a baseline method. 

\begin{figure}[b]
  \includegraphics[width=0.48\textwidth]{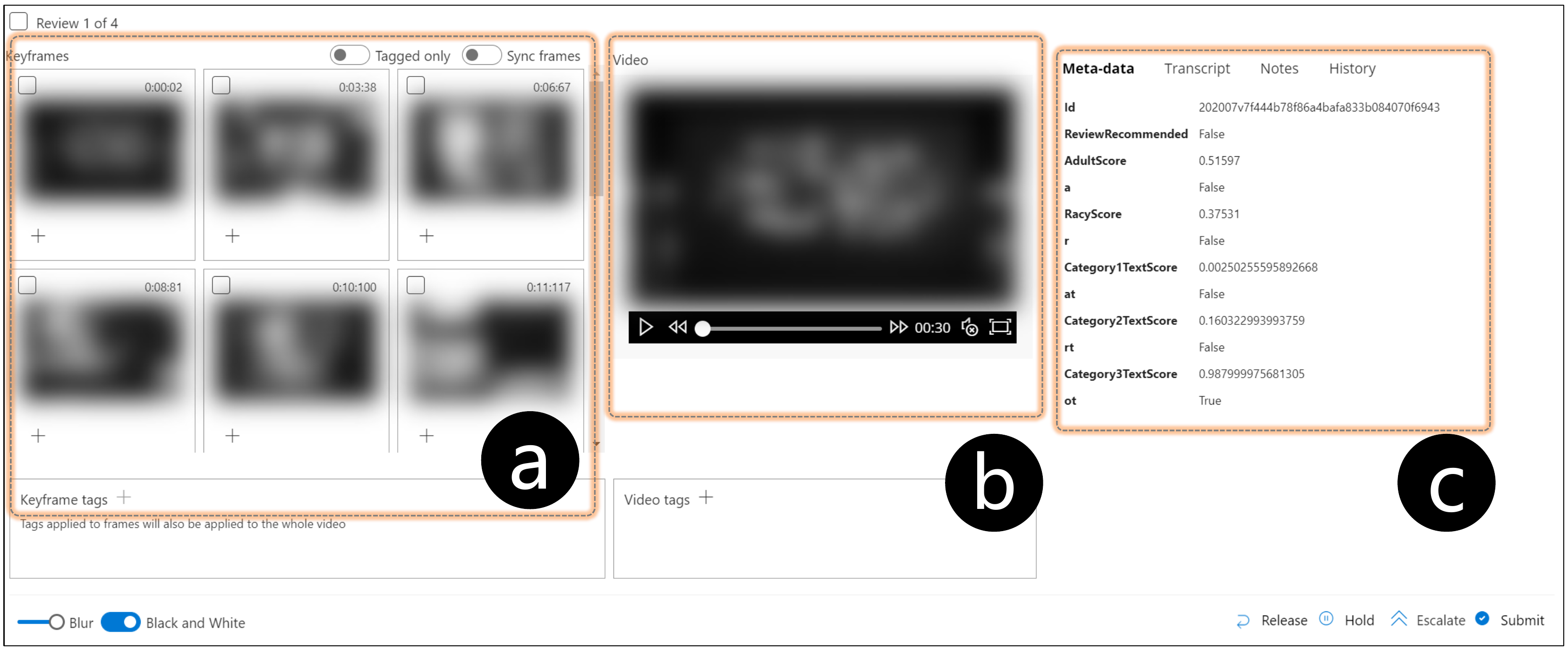}
  \caption{ContentModerator~\cite{MS} consists of (a) small multiples to visualize video frames, (b) a player to browse both video and audio content, and (c) a form view presenting meta-data and risk information.}
  \label{fig:baseline}
\end{figure}

\textbf{Baseline.}
VideoModerator is established on the basis of idea of showing moderators' additional but essential information, which can possibly increase their performance in reviewing videos.
Our hypothesis is that our visualization designs are effective, such that the benefits of exhibiting machine insights are higher than the cost of reading additional information.
To validate this hypothesis, we conducted a task-based user study that compared VideoModerator with a baseline tool that merely presented original video content.
Although many reviewing tools~\cite{MS,LionBridgeAI} satisfied our requirements, we selected ContentModerator~\cite{MS} as the baseline method due to its detailed documents and intuitive interface (Fig.~\ref{fig:baseline}).
We re-implemented ContentModerator, which comprised video and frame components, to make it suitable for our studies.
Specifically, we changed the video component to make it similar to our video view and presented the same frames in Fig.~\ref{fig:baseline}a.

\textbf{Participants.}
Our experiment used a within-subject design.
We recruited $12$ volunteers ($6$ males and $6$ females) to complete two groups of video moderation tasks by using VideoModerator and the baseline tool.
Their average age was $24.3$ years old (range=$21$-$28$ years).
All participants could read basic data visualizations, such as bar charts, and were knowledgeable about machine learning models.
They were from various backgrounds, including \textit{computer science}, \textit{social science}, and \textit{industry design}.
Two of them had been involved in jobs related to video moderation, while the others knew it from news or social media.
Each participant was paid $10\$$ for the $1$-h study.

\textbf{Data and Tasks.}
To simulate the real environment of video moderation, we first collected a bunch of videos that were detected as possibly deviant by machine learning models.
We divided the videos into three groups in accordance with the video length, namely, the short group ($<5$ min), the long group ($>30$ min), and the medium group (between two bounds).
The video length indicated the difficulty of reviewing videos, in which longer videos usually required more reviewing efforts.
To construct a diverse dataset, we randomly selected videos from the three groups and ensured that each group contained at least two selected videos.
In total, we obtained $20$ videos, whose average duration was $8.6$ min (ranging from $10$ s to $41$ min).
Given that these videos may be wrongly recalled, we determined their ground-truth labels by collaborating with the two experts who were also involved in our discussion about e-commerce moderation policies (Sec. 3).
Five out of the $20$ videos were normal videos, whereas the others were deviant videos.

Video moderation refers to the process in which moderators first review a video and then assign a label to the video.
We defined one \textit{normal} label and four \textit{deviant} labels that corresponded to the four moderation policies discussed in Sec. 3.
Our task required the participants to moderate videos by assigning them a \textit{normal} or \textit{risk} label.
We used two videos to demonstrate the two reviewing tools and enable the participants to practice.
We used the rest of the videos for formal testing and obtained $18$ tasks (\textbf{T1}-\textbf{T18}) in total.

\textbf{Procedure.}
The study consisted of three stages, namely, training ($20$ min), testing ($30$ min), and interviewing ($10$ min).
At the first stage, we demonstrated the basic concepts of video moderation and introduced the four policies that would be used to review videos.
Considering that the participants had various backgrounds, we trained them with concrete examples to remove the ambiguity and ensure that they had the same level of understandings about the policies.
The testing stage was further divided into two similar phases ($15$ min).
At each phase, the participants were first required to complete a practice task and then finish the formal tasks.
During the practice task, the participants could freely explore reviewing tools and ask for our assistance at any time, which were not allowed for the formal tasks.
We counterbalanced the order of two reviewing tools and tasks for a fair comparison.
After the participants completed the formal tasks with one tool, they were asked to give a rating on it by completing a subjective questionnaire.

\begin{figure}[t]
  \centering
  \includegraphics[width=0.48\textwidth]{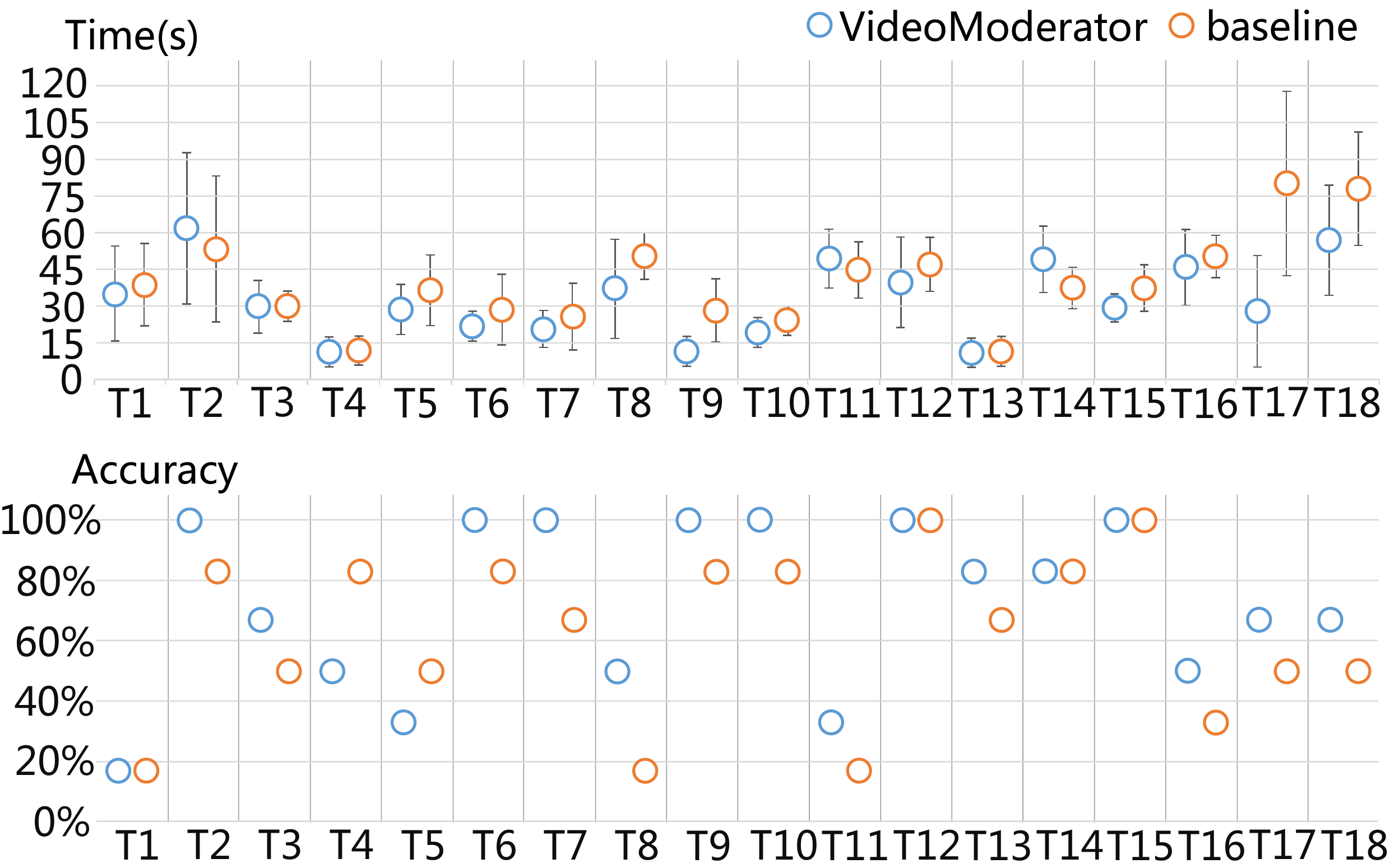}
  \caption{The completion time and task accuracy with $95\%$ CI.}
  \label{fig:study-results}
  \vspace{-6mm}
\end{figure}

\textbf{Questionnaire and Interview.}
To compare VideoModerator and the baseline tool further, we collected subjective ratings and qualitative feedback from the participants.
Specifically, the participants were required to give scores on each moderation tool by using five-point Likert scales from three perspectives, namely, \textit{easy of use}, \textit{easy of learning}, and \textit{effectiveness}.
We also interviewed the participants to understand their user experiences and \trevise{obtain their opinions about model interpretability}~\cite{Liu2017towards}.
The questions were as follows:
\begin{enumerate}[nosep]
  \renewcommand{\labelenumi}{\textbf{Q\theenumi}}
  \item \textit{How often each component was used when reviewing videos?}
  \item \textit{What order was followed when using these components?}
  \item \textit{What are your suggestions for further improvements?}
\end{enumerate}

\textbf{Results.}
We defined two crucial metrics, namely, \textit{time cost} and \textit{accuracy}, to measure the performance of each reviewing tool.
For each task, we calculated the average time cost of moderating the video and obtained the accuracy by comparing the labels assigned by the participants with the ground-truth labels.
In accordance with Fig.~\ref{fig:study-results}, the average time cost of our system is less than that of the baseline tool ($32.5 s<39.6 s$), but our accuracy is higher ($72\%>62\%$).
The results indicated that VideoModerator could achieve better performance than the baseline tool in terms of the two metrics.
We also observed three tasks, namely, \textbf{T2}, \textbf{T11}, and \textbf{T14}, in which the participants spent more time with VideoModerator.
Task T2 showed a game seller playing a $40$-min video game that contained diverse and numerous virtual scenes, which produced a large number of shot clusters.
The participants spent more time on browsing video frames with our tool but finally discovered deviant clips due to the machine-generated insights, which produced higher accuracy than the baseline method.
Task T4 contained a deviant clip at the end of the video, which was difficult to identify.
The participants who used the baseline tool promptly browsed the video but assigned a wrong label due to the neglect of the deviant clip.
Task T14 was a normal video but was detected as a deviant one by using machine learning models.
In general, the participants spent more time on thinking whether the machine-generated insights were correct by using VideoModerator.
The baseline tool did not present such contradicted information so that the participants could make quick decisions and achieve comparable performance in T14.
\trevise{
By comparing the statistics reported in the experiments, we found that there was a significant improvement over the machine learning models due to the benefits of human involvement.
Despite this comparison is not rigorous, the huge increase indicates that it is necessary to involve users into the moderation process by combining human and machine intelligence.
}

\begin{figure}[t]
  \centering
  \includegraphics[width=0.48\textwidth]{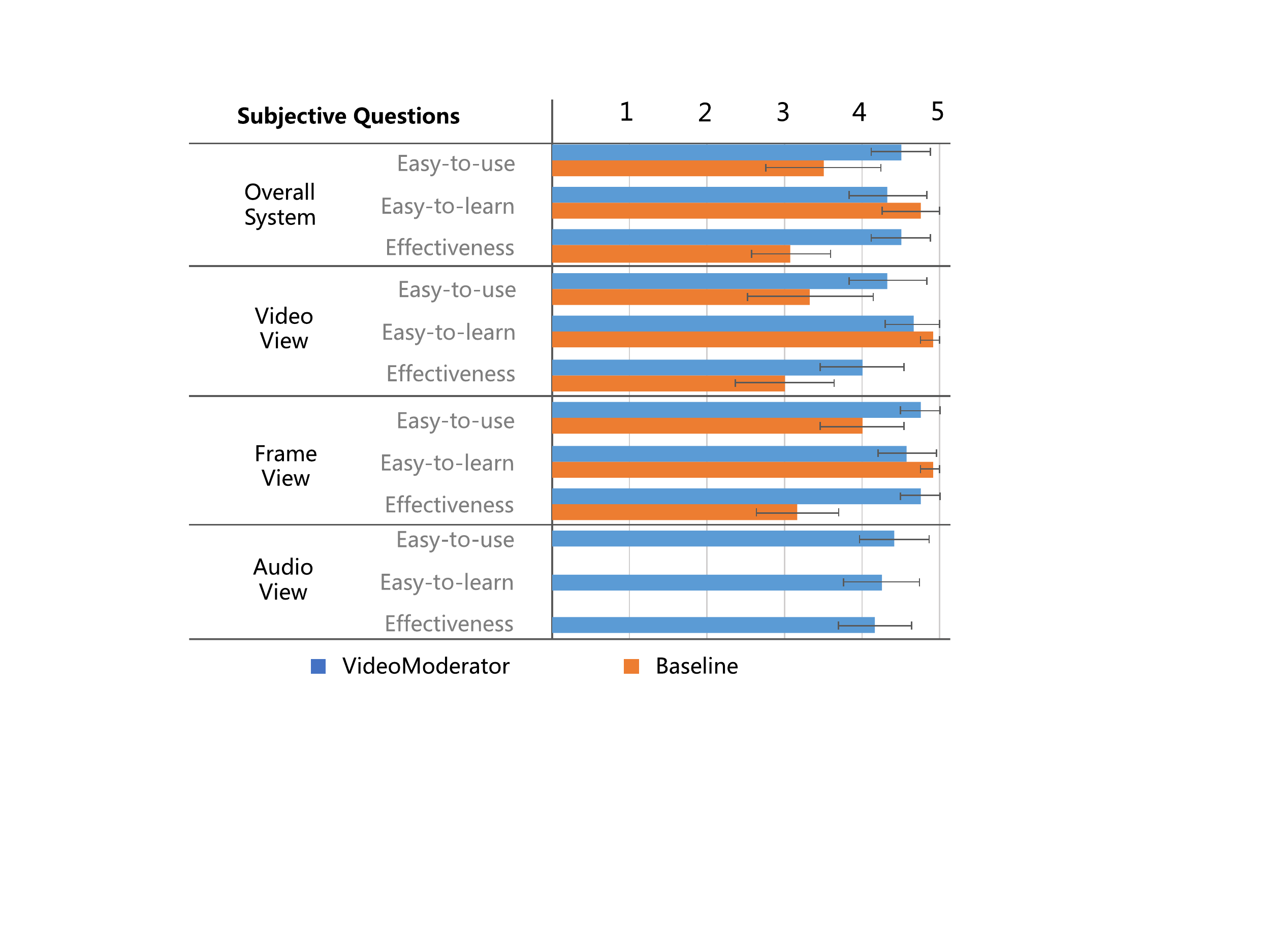}
  \caption{Analysis of the ratings: all ratings of VideoModerator were greater than four ($95\%$CI), which was very encouraging.}
  \label{fig:questionaire}
  \vspace{-6mm}
\end{figure}

The participants also filled questionnaires to compare the two reviewing tools with regard to \textit{easy of use}, \textit{easy of learning}, and \textit{effectiveness}.
Figure~\ref{fig:questionaire} presents the subjective questions and average user scores.
The two systems have comparable performance in terms of \textit{easy of use} and \textit{easy of learning}.
Considering that all scores are above $4$, we concluded that novice users could complete complex moderation tasks with minimum guidance by using VideoModerator.
Moreover, VideoModerator and its two components, namely, the video and frame views, are considered more effective than the baseline conditions.
This encouraging outcome indicates that the benefits of visualizing machine-generated insights are larger than the costs of integrating additional information.
Our audio view could also be considered useful due to its high scores.

\textbf{User Feedback.}
We collected the participants' qualitative feedback about VideoModerator to understand \trevise{model interpretability} and the usefulness of each component (Q1), discuss the general workflow of moderators (Q2), and obtain general suggestions (Q3).

\textit{Usability and Interpretability .}
Ten participants were impressed by the frame view that visually summarized video content and the rose diagrams that informed users of machine-generated risk labels.
P1 mentioned that``The scene projection and shot-clustering methods remarkably decreased the search space and saved the time in watching duplicate frames.''
\trevise{The participants confirmed the benefits of integrating risk information and the intuitiveness of the visualization designs.}
However, the segmented timeline was helpful but was only used when users could not locate risk frames or deviant audio clips.
P11 mentioned that ``I only use it to confirm whether the video is normal because the timeline already demonstrated the overall risk distribution.''
\trevise{We observed that the timeline seemed to be more isolated from the video content than the other visualization components.
We infer that users intend to receive the machine insights which are deeply embedded with the video context.}
The audio view was not as frequently used as the two other views because most deviant videos could be determined solely by using visual information.
However, the storyline-based design was appreciated by the participants due to its intuitiveness and expressiveness for visually summarizing audio content.
As P2 indicated, ``I could not listen to multiple audio clips at once, but, now, I could `watch' them because they were transformed into an image.''
\trevise{This comment indicates the big advantages of providing risk information once the machine-generated findings can be easily interpreted.}

\textit{Workflow.}
The participants showed diverse workflows because of their preferences and personal working experiences.
Eight participants started moderation tasks with the frame view, whereas two participants started with the audio view.
The rest of the participants did not follow any specific workflow and decided where to start in accordance with the shown risks.
P6 mentioned that``I intended to start with the frame view or audio view because they directly presented the detected deviant content.''
Despite that all participants confirmed that the video view was necessary, they thought it mainly served as a detailed view.
Most of them used the video view as the second step.
The audio view was usually used as the third step to inspect audio content.
The video view with the segmented timeline would be again to confirm whether the video is normal if no evident deviant clips were discovered.

\textit{Suggestions.}
The participants provided valuable suggestions on how to improve VideoModerator.
P10 suggested that ``It would be better to enable users to combine similar scenes so that users could focus on most risky scenes in the frame view.''
P8 provided a practical advice for the audio view that enabling users to highlight frames by clicking the horizontal bars could foster a better review of multimodal content.

%% file: sec7_conclusion.tex
\section{Discussion}
This section summarizes the generalization, lessons learned, limitations and future work for the proposed system.

\textbf{Generalization.}
Although our study is established on e-commerce livestreaming videos, the proposed system can be extended to other advertising videos because these videos share the same narrative elements.
In addition to e-commerce, VideoModerator can also be applied to other applications (e.g., surveillance) because the framework employs a ``learning with reviewing'' strategy that improves machine learning models when dealing with a new type of videos.
Video moderation is an inter-discipline among \textit{video summarization}, \textit{video visualization}~\cite{Chen2021VisCommentator}, and \textit{video annotation}~\cite{Hoferlin2012a}.
Generally, video moderation is regarded as a more complicated task than the video annotation task~\cite{Deng2021a} due to a larger search space that requires well-trained human moderators.
However, the shared workflow in which crowd workers first review videos and then assign labels makes it possible to extend our framework to video annotation.
Our storyline-based design sheds light on text visualizations.
Despite that we mainly focus on risk words, the storyline-based design can be easily extended to temporal text data, in which named entities can replace risk words to construct storylines.

\textbf{Lessons Learned.}
We present the design lessons learned when developing VideoModerator.
The visual exploration of machine-generated insights can foster an enhanced collaboration between models and human moderators.
Although the typical framework (Fig.~\ref{fig:mixed-initiative-framework}) has already integrated automatic models to moderate videos, it mainly targets at the machine side while ignoring the possibility of improving moderators' performance.
Existing moderation tools~\cite{MS,LionBridgeAI} merely present original videos; thus, the reviewing process is tedious and time consuming.
Our study successfully demonstrates that showing moderators additional necessary information will not hinder the reviewing processes but improve their time efficiency and accuracy.
Hence, our framework provides a new paradigm for video moderation systems.
\trevise{
We also summarize three design guidelines to inspire the usage of explainable artificial intelligence (AI) in visual analytic systems.
a) \textit{Interpretability:} as ``a picture is worth a thousand words,'' it becomes a common practice to visualize the machine-generated scores rather than telling users the exact numbers.
b) \textit{Proximity:} the visualizations of AI-generated results should be closely accompanied with the original information so that users can understand data insights promptly.
c) \textit{Simplicity:} it is necessary to avoid over-exaggerated design or enable excessive control because the cognitive capability of users may be overwhelmed.
}

\textbf{Limitations and future work.}
Our work has several limitations.
First, a gap exists between high-level risk labels and low-level risk words, which hinders the usage of the ``learning with reviewing'' strategy in ASR techniques.
To bridge the gap, we intend to detect risk words from audio clips directly so that the labeled videos could be used as the training data.
Second, the projection and clustering methods may produce numerous frame groups when dealing with videos that contain diverse scenes or shots.
We aim to adopt a hierarchical clustering method to combine redundant video shots.
Third, the scope of this study mainly focuses on moderating off-line videos.
Some online features of livestreamings, such as the interactions between livestreamers and audience, are also important to demonstrate underlying risks, which inspires promising future research.

\section{Conclusion}
We have presented VideoModerator, a risk-aware framework that enables quick navigation and supports fast moderation of deviant videos in e-commerce.
The framework bridges the gap between human moderators and machine learning models through interactive multimodal video visualizations.
First, our framework adopts a ``learning with reviewing'' strategy that iteratively improves model accuracy by using the ground-truth labels provided by users.
Second, we propose an interactive interface that visualizes multimodal machine-generated insights to foster an improved mixed-initiative environment.
Our evaluation indicates that VideoModerator can assist users in promptly understanding video content and effectively discovering deviant clips.
In the future, we plan to deploy our system in real-world applications and evaluate it through crowd-sourcing studies.

%% file: paper.bbl
\begin{thebibliography}{10}

\bibitem{Youtube}
{An introduction to YouTube policies and guidelines}.
\newblock Website, 2020.
\newblock Retrieved Dec. 1st, 2020 from
  \url{https://creatoracademy.youtube.com/page/lesson/copyright-guidelines#strategies-zippy-link-3}.

\bibitem{LionBridgeAI}
{Content Moderation Services}.
\newblock Website, 2020.
\newblock Retrieved Dec. 1st, 2020 from
  \url{https://lionbridge.ai/services/content-moderation/}.

\bibitem{MS}
{Video moderation with the Review tool}.
\newblock Website, 2020.
\newblock Retrieved Dec. 1st, 2020 from
  \url{https://docs.microsoft.com/en-us/azure/cognitive-services/content-moderator/video-moderation-human-review}.

\bibitem{Twitter}
Violent threats policy.
\newblock Website, 2020.
\newblock Retrieved Dec. 1st, 2020 from
  \url{https://help.twitter.com/en/rules-and-policies/violent-threats-glorification}.

\bibitem{Andrienko2021}
N.~Andrienko, G.~Andrienko, S.~Miksch, H.~Schumann, and S.~Wrobel.
\newblock A theoretical model for pattern discovery in visual analytics.
\newblock {\em Visual Informatics}, 5(1):23--33, 2021.

\bibitem{Nievas2011a}
E.~Bermejo~Nievas, O.~Deniz~Suarez, G.~Bueno~García, and R.~Sukthankar.
\newblock Violence {Detection} in {Video} {Using} {Computer} {Vision}
  {Techniques}.
\newblock In {\em Computer {Analysis} of {Images} and {Patterns}}, pp.
  332--339, 2011.

\bibitem{Borgo2012a}
R.~Borgo, M.~Chen, B.~Daubney, E.~Grundy, G.~Heidemann, B.~Höferlin,
  M.~Höferlin, H.~Leitte, D.~Weiskopf, and X.~Xie.
\newblock State of the {Art} {Report} on {Video}-{Based} {Graphics} and {Video}
  {Visualization}.
\newblock {\em Computer Graphics Forum}, 31(8):2450--2477, 2012.

\bibitem{Botchen2008a}
R.~P. Botchen, S.~Bachthaler, F.~Schick, M.~Chen, G.~Mori, D.~Weiskopf, and
  T.~Ertl.
\newblock Action-{Based} {Multifield} {Video} {Visualization}.
\newblock {\em IEEE Transactions on Visualization and Computer Graphics},
  14(4):885--899, 2008.

\bibitem{Cai2019a}
J.~Cai and D.~Y. Wohn.
\newblock {Live Streaming Commerce: Uses and Gratifications Approach to
  Understanding Consumers’ Motivations}.
\newblock In {\em Proceedings of the Hawaii International Conference on System
  Sciences}, 2019.

\bibitem{Chancellor2017a}
S.~Chancellor, Y.~Kalantidis, J.~A. Pater, M.~De~Choudhury, and D.~A. Shamma.
\newblock Multimodal {Classification} of {Moderated} {Online} {Pro}-{Eating}
  {Disorder} {Content}.
\newblock In {\em Proceedings of the {ACM CHI} {Conference} on {Human}
  {Factors} in {Computing} {Systems}}, pp. 3213--3226, May 2017.

\bibitem{Chen2006a}
M.~Chen, R.~Botchen, R.~Hashim, D.~Weiskopf, T.~Ertl, and I.~Thornton.
\newblock Visual {Signatures} in {Video} {Visualization}.
\newblock {\em IEEE Transactions on Visualization and Computer Graphics},
  12(5):1093--1100, 2006.

\bibitem{Chen2021VisCommentator}
Z.~Chen, S.~Ye, X.~Chu, H.~Xia, H.~Zhang, H.~Qu, and Y.~Wu.
\newblock {Augmenting Sports Videos with VisCommentator}.
\newblock {\em IEEE Transactions on Visualization and Computer Graphics}, p. to
  appear, 2022.

\bibitem{Chinchor2010a}
N.~A. Chinchor, J.~J. Thomas, P.~C. Wong, M.~G. Christel, and W.~Ribarsky.
\newblock Multimedia {Analysis} + {Visual} {Analytics} = {Multimedia}
  {Analytics}.
\newblock {\em IEEE Computer Graphics and Applications}, 30(5):52--60, 2010.

\bibitem{Chollet2018a}
F.~Chollet.
\newblock {The Limitations of Deep Learning}.
\newblock In {\em Deep learning with Python}. Manning New York, 2018.

\bibitem{Cox2008a}
M.~A. Cox and T.~F. Cox.
\newblock {Multidimensional Scaling}.
\newblock In {\em Handbook of Data Visualization}, pp. 315--347. Springer,
  2008.

\bibitem{Daniel2003a}
G.~Daniel and {Min Chen}.
\newblock Video {V}isualization.
\newblock In {\em Proceedings of the {IEEE} {Visualization}}, pp. 409--416,
  2003.

\bibitem{Deng2021a}
D.~Deng, J.~Wu, J.~Wang, Y.~Wu, X.~Xie, Z.~Zhou, H.~Zhang, X.~L. Zhang, and
  Y.~Wu.
\newblock {EventAnchor: Reducing Human Interactions in Event Annotation of
  Racket Sports Videos}.
\newblock In {\em Proceedings of the ACM CHI Conference on Human Factors in
  Computing Systems}, pp. 1--13, 2021.

\bibitem{Duffy2015a}
B.~Duffy, J.~Thiyagalingam, S.~Walton, D.~J. Smith, A.~Trefethen, J.~C.
  Kirkman-Brown, E.~A. Gaffney, and M.~Chen.
\newblock Glyph-{Based} {Video} {Visualization} for {Semen} {Analysis}.
\newblock {\em IEEE Transactions on Visualization and Computer Graphics},
  21(8):980--993, 2015.

\bibitem{Filipov2021Circles}
V.~Filipov, V.~Schetinger, K.~Raminger, N.~Soursos, S.~Zapke, and S.~Miksch.
\newblock Gone full circle: A radial approach to visualize event-based networks
  in digital humanities.
\newblock {\em Visual Informatics}, 5(1):45--60, 2021.

\bibitem{Giannakopoulos2010a}
T.~Giannakopoulos, A.~Makris, D.~Kosmopoulos, S.~Perantonis, and
  S.~Theodoridis.
\newblock Audio-{Visual} {Fusion} for {Detecting} {Violent} {Scenes} in
  {Videos}.
\newblock In {\em Artificial {Intelligence}: {Theories}, {Models} and
  {Applications}}, pp. 91--100, 2010.

\bibitem{Gillespie2018a}
T.~Gillespie.
\newblock {\em Custodians of the Internet: Platforms, content moderation, and
  the hidden decisions that shape social media}.
\newblock Yale University Press, 2018.

\bibitem{Guo2021DanceVis}
H.~Guo, S.~Zou, Y.~Xu, H.~Yang, J.~Wang, H.~Zhang, and W.~Chen.
\newblock {DanceVis: Toward Better Understanding of the Cheer and Dance
  Training}.
\newblock {\em Journal of Visualization}, p. to appear, 2021.

\bibitem{Gupta2018a}
D.~Gupta, I.~Sen, N.~Sachdeva, P.~Kumaraguru, and A.~B. Buduru.
\newblock Empowering {First} {Responders} through {Automated} {Multimodal}
  {Content} {Moderation}.
\newblock In {\em Proceedings of the {IEEE} {International} {Conference} on
  {Cognitive} {Computing}}, pp. 1--8, July 2018.

\bibitem{Gygli2014a}
M.~Gygli, H.~Grabner, H.~Riemenschneider, and L.~Van~Gool.
\newblock Creating {Summaries} from {User} {Videos}.
\newblock In {\em Proceedings of the {European} {Conference} on {Computer}
  {Vision}}, pp. 505--520, 2014.

\bibitem{Han2021NetV}
D.~Han, J.~Pan, X.~Zhao, and W.~Chen.
\newblock Netv.js: A web-based library for high-efficiency visualization of
  large-scale graphs and networks.
\newblock {\em Visual Informatics}, 5(1):61--66, 2021.

\bibitem{Hanson2019a}
A.~Hanson, K.~Pnvr, S.~Krishnagopal, and L.~Davis.
\newblock Bidirectional {Convolutional} {LSTM} for the {Detection} of
  {Violence} in {Videos}.
\newblock In {\em Proceedings of the European Conference on Computer Vision
  Workshops}, pp. 280--295, 2019.

\bibitem{ColorBrewer}
M.~Harrower and C.~A. Brewer.
\newblock {ColorBrewer.org: An Online Tool for Selecting Colour Schemes for
  Maps}.
\newblock {\em The Cartographic Journal}, 40:27--37, 2003.

\bibitem{Hassner2012a}
T.~Hassner, Y.~Itcher, and O.~Kliper-Gross.
\newblock Violent flows: {Real}-time detection of violent crowd behavior.
\newblock In {\em Proceedings of the {IEEE} {Conference} on {Computer} {Vision}
  and {Pattern} {Recognition} {Workshops}}, pp. 1--6, June 2012.

\bibitem{Haubold2005a}
A.~Haubold and J.~R. Kender.
\newblock Augmented segmentation and visualization for presentation videos.
\newblock In {\em Proceedings of the {ACM} international conference on
  {Multimedia}}, pp. 51--60, 2005.

\bibitem{He2016a}
K.~He, X.~Zhang, S.~Ren, and J.~Sun.
\newblock {Deep Residual Learning for Image Recognition}.
\newblock In {\em Proceedings of the IEEE Conference on Computer Vision and
  Pattern Recognition}, pp. 770--778, 2016.

\bibitem{Hoferlin2012a}
B.~Hoferlin, R.~Netzel, M.~Hoferlin, D.~Weiskopf, and G.~Heidemann.
\newblock {Inter-active Learning of Ad-hoc Classifiers for Video Visual
  Analytics}.
\newblock In {\em Proceedings of the {IEEE} {Conference} on {Visual}
  {Analytics} {Science} and {Technology}}, pp. 23--32, 2012.

\bibitem{Hoferlin2013a}
M.~Hoferlin, B.~Hoferlin, G.~Heidemann, and D.~Weiskopf.
\newblock Interactive {Schematic} {Summaries} for {Faceted} {Exploration} of
  {Surveillance} {Video}.
\newblock {\em IEEE Transactions on Multimedia}, 15(4):908--920, 2013.

\bibitem{Hoferlin2015a}
B.~Höferlin, M.~Höferlin, G.~Heidemann, and D.~Weiskopf.
\newblock {Scalable Video Visual Analytics}.
\newblock {\em Information Visualization}, 14(1):10--26, 2015.

\bibitem{Jamonnak2021Geo}
S.~Jamonnak, Y.~Zhao, X.~Huang, and M.~Amiruzzaman.
\newblock {Geo-Context Aware Study of Vision-Based Autonomous Driving Models
  and Spatial Video Data}.
\newblock {\em IEEE Transactions on Visualization and Computer Graphics}, p. to
  appear, 2022.

\bibitem{Jhaver2018a}
S.~Jhaver, S.~Ghoshal, A.~Bruckman, and E.~Gilbert.
\newblock {Online Harassment and Content Moderation: The Case of Blocklists}.
\newblock {\em ACM Transactions on Computer-Human Interaction}, 25(2):1--33,
  2018.

\bibitem{John2019a}
M.~John, K.~Kurzhals, and T.~Ertl.
\newblock Visual {Exploration} of {Topics} in {Multimedia} {News} {Corpora}.
\newblock In {\em Proceedings of the {International} {Conference} {Information}
  {Visualisation}}, pp. 241--248, 2019.

\bibitem{Jackle2016a}
D.~{Jäckle}, F.~{Fischer}, T.~{Schreck}, and D.~A. {Keim}.
\newblock Temporal mds plots for analysis of multivariate data.
\newblock {\em IEEE Transactions on Visualization and Computer Graphics},
  22(1):141--150, 2016.

\bibitem{Kang2006a}
H.-W. Kang.
\newblock Space-{Time} {Video} {Montage}.
\newblock In {\em Proceedings of the {IEEE} {Computer} {Society} {Conference}
  on {Computer} {Vision} and {Pattern} {Recognition}}, 2006.

\bibitem{Kiesler2012a}
S.~Kiesler, R.~Kraut, P.~Resnick, and A.~Kittur.
\newblock {Regulating Behavior in Online Communities}.
\newblock In {\em Building successful online communities: Evidence-based social
  design}, pp. 125--178, 2012.

\bibitem{Kurzhals2016a}
K.~Kurzhals, M.~John, F.~Heimerl, P.~Kuznecov, and D.~Weiskopf.
\newblock Visual {Movie} {Analytics}.
\newblock {\em IEEE Transactions on Multimedia}, 18(11):2149--2160, 2016.

\bibitem{Lampe2004a}
C.~Lampe and P.~Resnick.
\newblock {Slash (dot) and Burn: Distributed Moderation in a Large Online
  Conversation Space}.
\newblock In {\em Proceedings of the SIGCHI Conference on Human Factors in
  Computing Systems}, pp. 543--550, 2004.

\bibitem{Lan2021RallyComparator}
J.~Lan, J.~Wang, X.~Shu, Z.~Zhou, H.~Zhang, and Y.~Wu.
\newblock {RallyComparator: Visual Comparison of the Multivariate and Spatial
  Stroke Sequence in A Table Tennis Rally}.
\newblock {\em Journal of Visualization}, p. to appear, 2021.

\bibitem{Liu2017towards}
M.~Liu, J.~Shi, Z.~Li, C.~Li, J.~Zhu, and S.~Liu.
\newblock Towards better analysis of deep convolutional neural networks.
\newblock {\em IEEE Transactions on Visualization and Computer Graphics},
  23(1):91--100, 2017.

\bibitem{Ma2020a}
C.-X. Ma, J.-C. Song, Q.~Zhu, K.~Maher, Z.-Y. Huang, and H.-A. Wang.
\newblock {EmotionMap}: {Visual} {Analysis} of {Video} {Emotional} {Content} on
  a {Map}.
\newblock {\em Journal of Computer Science and Technology}, 35(3):576--591,
  2020.

\bibitem{Hutchison2013a}
K.~M. Mahmoud, M.~A. Ismail, and N.~M. Ghanem.
\newblock {VSCAN}: {An} {Enhanced} {Video} {Summarization} {Using}
  {Density}-{Based} {Spatial} {Clustering}.
\newblock In {\em Proceedings of the {International} {Conference} on {Image}
  {Analysis} and {Processing}}, pp. 733--742, 2013.

\bibitem{Meghdadi2013a}
A.~H. Meghdadi and P.~Irani.
\newblock Interactive {Exploration} of {Surveillance} {Video} through {Action}
  {Shot} {Summarization} and {Trajectory} {Visualization}.
\newblock {\em IEEE Transactions on Visualization and Computer Graphics},
  19(12):2119--2128, 2013.

\bibitem{Mnih2015a}
V.~Mnih, K.~Kavukcuoglu, D.~Silver, A.~A. Rusu, J.~Veness, M.~G. Bellemare,
  A.~Graves, M.~Riedmiller, A.~K. Fidjeland, G.~Ostrovski, et~al.
\newblock Human-level control through deep reinforcement learning.
\newblock {\em Nature}, 518(7540):529--533, 2015.

\bibitem{Moritz2020a}
N.~Moritz, T.~Hori, and J.~Le.
\newblock Streaming automatic speech recognition with the transformer model.
\newblock In {\em Proceedings of the IEEE International Conference on
  Acoustics, Speech and Signal Processing}, pp. 6074--6078. IEEE, 2020.

\bibitem{Nguyen2012a}
C.~Nguyen, Y.~Niu, and F.~Liu.
\newblock Video summagator: an interface for video summarization and
  navigation.
\newblock In {\em Proceedings of the {SIGCHI} {Conference} on {Human} {Factors}
  in {Computing} {Systems}}, pp. 647--650, 2012.

\bibitem{Peixoto2019a}
B.~Peixoto, B.~Lavi, J.~P. Pereira~Martin, S.~Avila, Z.~Dias, and A.~Rocha.
\newblock Toward {Subjective} {Violence} {Detection} in {Videos}.
\newblock In {\em Proceedings of the {IEEE} {International} {Conference} on
  {Acoustics}, {Speech} and {Signal} {Processing}}, pp. 8276--8280, May 2019.

\bibitem{Poco2017a}
J.~Poco and J.~Heer.
\newblock {Reverse-Engineering Visualizations: Recovering Visual Encodings from
  Chart Images}.
\newblock {\em Computer Graphics Forum}, 36(3):353--363, 2017.

\bibitem{Ren2017a}
S.~Ren, K.~He, R.~Girshick, and J.~Sun.
\newblock Faster {R}-{CNN}: {Towards} {Real}-{Time} {Object} {Detection} with
  {Region} {Proposal} {Networks}.
\newblock {\em IEEE Transactions on Pattern Analysis and Machine Intelligence},
  39(6):1137--1149, 2017.

\bibitem{Renoust2017a}
B.~Renoust, H.~Ren, G.~Melançon, M.-L. Viaud, and S.~Satoh.
\newblock {FaceCloud}: {Heterogeneous} {Cloud} {Visualization} of {Multiplex}
  {Networks} for {Multimedia} {Archive} {Exploration}.
\newblock In {\em Proceedings of the {ACM} International Conference on
  {Multimedia}}, pp. 1235--1236, 2017.

\bibitem{Roberts2016a}
S.~T. Roberts.
\newblock {Commercial Content Moderation: Digital Laborers' Dirty Work}.
\newblock In {\em Media Studies Publications}, 2016.

\bibitem{Romero2008a}
M.~Romero, J.~Summet, J.~Stasko, and G.~Abowd.
\newblock Viz-{A}-{Vis}: {Toward} {Visualizing} {Video} through {Computer}
  {Vision}.
\newblock {\em IEEE Transactions on Visualization and Computer Graphics},
  14(6):1261--1268, 2008.

\bibitem{Singh2018a}
A.~Singh, D.~Patil, and S.~Omkar.
\newblock Eye in the {Sky}: {Real}-{Time} {Drone} {Surveillance} {System}
  ({DSS}) for {Violent} {Individuals} {Identification} {Using} {ScatterNet}
  {Hybrid} {Deep} {Learning} {Network}.
\newblock In {\em Proceedings of the {IEEE} {Conference} on {Computer} {Vision}
  and {Pattern} {Recognition} {Workshops}}, pp. 1742--1750, June 2018.

\bibitem{Soure2021CoUX}
E.~J. Soure, E.~Kuang, M.~Fan, and J.~Zhao.
\newblock {CoUX: Collaborative Visual Analysis of Think-Aloud Usability Test
  Videos for Digital Interfaces}.
\newblock {\em IEEE Transactions on Visualization and Computer Graphics}, p. to
  appear, 2022.

\bibitem{Sun2019a}
Y.~Sun, X.~Shao, X.~Li, Y.~Guo, and K.~Nie.
\newblock {How Live Streaming Influences Purchase Intentions in Social
  Commerce: An IT Affordance Perspective}.
\newblock {\em Electronic Commerce Research and Applications}, 37:100886, 08
  2019.

\bibitem{Szegedy2016a}
C.~Szegedy, V.~Vanhoucke, S.~Ioffe, J.~Shlens, and Z.~Wojna.
\newblock Rethinking the {Inception} {Architecture} for {Computer} {Vision}.
\newblock In {\em Proceedings of the {IEEE} {Conference} on {Computer} {Vision}
  and {Pattern} {Recognition}}, pp. 2818--2826, 2016.

\bibitem{Tang2020a}
T.~Tang, R.~Li, X.~Wu, S.~Liu, J.~Knittel, S.~Koch, L.~Yu, P.~Ren, T.~Ertl, and
  Y.~Wu.
\newblock {Plotthread: Creating Expressive Storyline Visualizations using
  Reinforcement Learning}.
\newblock {\em IEEE Transactions on Visualization and Computer Graphics},
  27(2):294--303, 2021.

\bibitem{Tang2018a}
T.~Tang, S.~Rubab, J.~Lai, W.~Cui, L.~Yu, and Y.~Wu.
\newblock {iStoryline: Effective Convergence to Hand-drawn Storylines}.
\newblock {\em IEEE Transactions on Visualization and Computer Graphics},
  25(1):769--778, 2018.

\bibitem{Van2008a}
L.~Van~der Maaten and G.~Hinton.
\newblock Visualizing data using t-sne.
\newblock {\em Journal of machine learning research}, 9(11):2579--2605, 2008.

\bibitem{Wang2021Perception}
J.~Wang, X.~Cai, J.~Su, Y.~Liao, and Y.~Wu.
\newblock {What Makes a Scatterplot Hard to Comprehend: Data Size and Pattern
  Salience Matter}.
\newblock {\em Journal of Visualization}, p. to appear, 2021.

\bibitem{Wang2021Tac}
J.~Wang, J.~Wu, A.~Cao, Z.~Zhou, H.~Zhang, and Y.~Wu.
\newblock {Tac-Miner: Visual Tactic Mining for Multiple Table Tennis Matches}.
\newblock {\em {IEEE} Transactions on Visualization and Computer Graphics},
  27(6):2770--2782, 2021.

\bibitem{Di2021Bus}
D.~Weng, C.~Zheng, Z.~Deng, M.~Ma, J.~Bao, Y.~Zheng, M.~Xu, and Y.~Wu.
\newblock {Towards Better Bus Networks: A Visual Analytics Approach}.
\newblock {\em {IEEE} Transactions on Visualization and Computer Graphics},
  27(2):817--827, 2021.

\bibitem{Wu2020b}
A.~Wu and H.~Qu.
\newblock Multimodal {Analysis} of {Video} {Collections}: {Visual}
  {Exploration} of {Presentation} {Techniques} in {TED} {Talks}.
\newblock {\em IEEE Transactions on Visualization and Computer Graphics},
  26(7):2429 -- 2442, 2020.

\bibitem{Wu2021survey}
A.~Wu, Y.~Wang, X.~Shu, D.~Moritz, W.~Cui, H.~Zhang, D.~Zhang, and H.~Qu.
\newblock {AI4VIS: Survey on Artificial Intelligence Approaches for Data
  Visualization}.
\newblock {\em arXiv preprint arXiv:2102.01330}, 2021.

\bibitem{Wu2020a}
P.~Wu, J.~Liu, Y.~Shi, Y.~Sun, F.~Shao, Z.~Wu, and Z.~Yang.
\newblock Not only {Look}, {But} {Also} {Listen}: {Learning} {Multimodal}
  {Violence} {Detection} {Under} {Weak} {Supervision}.
\newblock In {\em Proceedings of the European Conference on Computer Vision},
  pp. 322--339, 2020.

\bibitem{Wu2021Co}
Y.~Wu, D.~Weng, Z.~Deng, J.~Bao, M.~Xu, Z.~Wang, Y.~Zheng, Z.~Ding, and
  W.~Chen.
\newblock {Towards Better Detection and Analysis of Massive Spatiotemporal
  Co-Occurrence Patterns}.
\newblock {\em IEEE Transactions on Intelligent Transportation Systems},
  22(6):3387--3402, 2021.

\bibitem{Xie2018a}
X.~Xie, X.~Cai, J.~Zhou, N.~Cao, and Y.~Wu.
\newblock A semantic-based method for visualizing large image collections.
\newblock {\em IEEE Transactions on Visualization and Computer Graphics},
  25(7):2362--2377, 2018.

\bibitem{Xie2021PassVizor}
X.~Xie, J.~Wang, H.~Liang, D.~Deng, S.~Cheng, H.~Zhang, W.~Chen, and Y.~Wu.
\newblock {PassVizor: Toward Better Understanding of the Dynamics of Soccer
  Passes}.
\newblock {\em {IEEE} Transactions on Visualization and Computer Graphics},
  27(2):1322--1331, 2021.

\bibitem{Ye2021Shuttle}
S.~Ye, Z.~Chen, X.~Chu, Y.~Wang, S.~Fu, L.~Shen, K.~Zhou, and Y.~Wu.
\newblock {ShuttleSpace: Exploring and Analyzing Movement Trajectory in
  Immersive Visualization}.
\newblock {\em {IEEE} Transactions on Visualization and Computer Graphics},
  27(2):860--869, 2021.

\bibitem{Ying2006a}
{Ying Li}, {Shih-Hung Lee}, {Chia-Hung Yeh}, and C.-C. Kuo.
\newblock Techniques for movie content analysis and skimming: tutorial and
  overview on video abstraction techniques.
\newblock {\em IEEE Signal Processing Magazine}, 23(2):79--89, 2006.

\bibitem{Yuan2021survey}
J.~Yuan, C.~Chen, W.~Yang, M.~Liu, J.~Xia, and S.~Liu.
\newblock {A Survey of Visual Analytics Techniques for Machine Learning}.
\newblock {\em Computational Visual Media}, 7(1):3--36, 2021.

\bibitem{Zahalka2014a}
J.~Zahalka and M.~Worring.
\newblock Towards interactive, intelligent, and integrated multimedia
  analytics.
\newblock In {\em Proceedings of the {IEEE} {Conference} on {Visual}
  {Analytics} {Science} and {Technology}}, pp. 3--12, 2014.

\bibitem{Zeng2020b}
H.~Zeng, X.~Shu, Y.~Wang, Y.~Wang, L.~Zhang, T.-C. Pong, and H.~Qu.
\newblock {EmotionCues}: {Emotion}-{Oriented} {Visual} {Summarization} of
  {Classroom} {Videos}.
\newblock {\em IEEE Transactions on Visualization and Computer Graphics}, p. to
  appear, 2020.

\bibitem{Zeng2020a}
H.~Zeng, X.~Wang, A.~Wu, Y.~Wang, Q.~Li, A.~Endert, and H.~Qu.
\newblock {EmoCo}: {Visual} {Analysis} of {Emotion} {Coherence} in
  {Presentation} {Videos}.
\newblock {\em IEEE Transactions on Visualization and Computer Graphics},
  26(1):927--937, 2020.

\bibitem{Zhao2021a}
Y.~Zhao, H.~Jiang, Q.~Chen, Y.~Qin, H.~Xie, Y.~Wu, S.~Liu, Z.~Zhou, J.~Xia, and
  F.~Zhou.
\newblock {Preserving Minority Structures in Graph Sampling}.
\newblock {\em IEEE Transactions on Visualization and Computer Graphics},
  27(2):1698--1708, 2021.

\end{thebibliography}
